\documentclass[useAMS,usenatbib]{mn2e}
\usepackage{epsfig}
\usepackage{epstopdf}
\usepackage{graphicx,amssymb,supertabular,longtable,lscape} 
\usepackage{amsmath}
\usepackage{textcomp}
\usepackage[T1]{fontenc}
\usepackage{times}

\def\apj{\mbox{ApJ}}
\def\apjl{\mbox{ApJL}}
\def\apjs{\mbox{ApJS}}

\def\mnras{\mbox{MNRAS}}
\def\aj{\mbox{AJ}}

\def\nat{\mbox{Nature}}
\def\aap{\mbox{A\&A}}

\def\mum{\textmu m }
\def\mums{\textmu m}

\title[AGN and star formation in HerMES IRS sources]
{The roles of star formation and AGN activity of IRS sources in the HerMES fields}

\author[A.~Feltre et al.]
{\parbox{\textwidth}{\raggedright A.~Feltre,$^{1,2}$\thanks{E-mail: \texttt{afeltre@eso.org}}
E.~Hatziminaoglou,$^{2}$
A. ~Hern\'an-Caballero,$^{3}$
J. ~Fritz,$^{4}$
A.~Franceschini,$^{1}$
J.~Bock,$^{5,6}$
A.~Cooray,$^{7,5}$
D.~Farrah,$^{8}$
E.A.~Gonz\'alez~Solares,$^{9}$
E.~Ibar,$^{10,11}$
K.G.~Isaak,$^{12}$
B.~Lo~Faro,$^{1}$
L.~Marchetti,$^{13,1}$
S.J.~Oliver,$^{14}$
M.J.~Page,$^{15}$
D.~Rigopoulou,$^{16,17}$
I.G.~Roseboom,$^{14,18}$
M.~Symeonidis$^{15}$ and
M.~Vaccari$^{1,19}$}\vspace{0.4cm}\\
\parbox{\textwidth}{\raggedright $^{1}$Dipartimento di Fisica e Astronomia, Universit\`{a} di Padova, vicolo Osservatorio, 3, 35122 Padova, Italy\\
$^{2}$ESO, Karl-Schwarzschild-Str. 2, 85748 Garching bei M\"unchen, Germany\\
$^{3}$Instituto de Fisica de Cantabria, CSIC-UC, Avenida de los Castros s/n, 39005, Santander, Spain\\
$^{4}$Sterrenkundig Observatorium, Vakgroep Fysica en Sterrenkunde Universeit, Gent, Krijgslaan 281, S9  9000 Gent\\
$^{5}$California Institute of Technology, 1200 E. California Blvd., Pasadena, CA 91125, USA\\
$^{6}$Jet Propulsion Laboratory, 4800 Oak Grove Drive, Pasadena, CA 91109, USA\\
$^{7}$Dept. of Physics \& Astronomy, University of California, Irvine, CA 92697, USA\\
$^{8}$Department of Physics, Virginia Tech, Blacksburg, VA 24061, USA\\
$^{9}$Institute of Astronomy, University of Cambridge, Madingley Road, Cambridge CB3 0HA, UK\\
$^{10}$UK Astronomy Technology Centre, Royal Observatory, Blackford Hill, Edinburgh EH9 3HJ, UK\\
$^{11}$Pontificia Universidad Cat\'olica de Chile, Departamento de Astronom\'ia y Astrof\'isica, Vicu\~na Mackenna 4860, Casilla 306, Santiago 22, Chile\\
$^{12}$ESA Research and Scientific Support Department, ESTEC/SRE-SA, Keplerlaan 1, 2201 AZ Noordwijk, The Netherlands\\
$^{13}$Department of Physical Sciences, The Open University, Milton Keynes MK7 6AA, UK\\
$^{14}$Astronomy Centre, Dept. of Physics \& Astronomy, University of Sussex, Brighton BN1 9QH, UK\\
$^{15}$Mullard Space Science Laboratory, University College London, Holmbury St. Mary, Dorking, Surrey RH5 6NT, UK\\
$^{16}$RAL Space, Rutherford Appleton Laboratory, Chilton, Didcot, Oxfordshire OX11 0QX, UK\\
$^{17}$Department of Astrophysics, Denys Wilkinson Building, University of Oxford, Keble Road, Oxford OX1 3RH, UK\\
$^{18}$Institute for Astronomy, University of Edinburgh, Royal Observatory, Blackford Hill, Edinburgh EH9 3HJ, UK\\
$^{19}$Astrophysics Group, Physics Department, University of the Western Cape, Private Bag X17, 7535, Bellville, Cape Town, South Africa}}

\begin{document}

\pagerange{\pageref{firstpage}--\pageref{lastpage}} \pubyear{2012}

\maketitle

\label{firstpage}

\begin{abstract}

In this work we explore the impact of the presence of an active galactic nucleus (AGN) on the mid- and far-infrared (IR) properties of galaxies as well as the effects of simultaneous AGN and starburst activity in these same galaxies.  To do this we apply a multi-component, multi-band spectral synthesis technique to a sample of 250 \mum selected galaxies of the {\it Herschel} Multi-tiered Extragalactic Survey (HerMES), with IRS spectra available for all galaxies. Our results confirm that the inclusion of the IRS spectra plays a crucial role in the spectral analysis of galaxies with an AGN component improving the selection of the best-fit hot dust (torus) model.

We find a correlation between the obscured star formation rate (SFR) derived from the IR luminosity of the starburst component, SFR$_{\rm IR}$ and SFR$_{\rm PAH}$, derived from the luminosity of the PAH features, L$_{\rm PAH}$, with SFR$_{\rm FIR}$ taking higher values than SFR$_{\rm PAH}$. The correlation is different for AGN- and starburst-dominated objects. The ratio of L$_{\rm PAH}$ to that of the starburst component, L$_{\rm PAH}$/L$_{\rm SB}$, is  almost constant for AGN-dominated objects but decreases with increasing L$_{\rm SB}$ for starburst-dominated objects. SFR$_{\rm FIR}$ increases with the accretion luminosity, L$_{\rm acc}$, with the increase less prominent for the very brightest, unobscured AGN-dominated sources. 

We find no correlation between the masses of the hot (AGN-heated) and cold (starburst-heated) dust components. We interpret this as a non-constant fraction of gas driven by the gravitational effects to the AGN while the starburst is ongoing. We also find no evidence of the AGN affecting the temperature of the cold dust component, though this conclusion is mostly based on objects with a non-dominant AGN component. We conclude that our findings do not provide evidence that the presence of AGN affects the star formation process in the host galaxy, but rather that the two phenomena occur simultaneously over a wide range of luminosities.

\end{abstract}


\section{Introduction}\label{intro}
There is now both observational and theoretical evidence for an intimate link between the growth of galaxies and the supermassive black holes (SMBHs) residing at their centres. On the theoretical front, cosmological simulations and semi-analytic models find it necessary to include feedback from Active Galactic Nuclei (AGN) to suppress star formation in massive galaxies \citep[e.g.][]{bower06,booth09,croton06} in order to account for the observed galaxy mass function simultaneously at low and high redshifts. Observations have shown that the masses of SMBHs, and those of the bulges in galaxies, follow a tight proportionality \citep[][]{magorrian98,ferrarese00,tremaine02}, and that the peak of the quasar number density \citep[e.g.][]{boyle98,richards06} coincides with that of the star formation history of the universe \citep[e.g.][]{madau98,heavens04}. Moreover, AGN and circumnuclear star formation are often found to co-exist in galaxies at all redshifts \citep[e.g.][]{farrah03,alexander05}, usually in systems with significant dust obscuration, and some authors have claimed evidence for direct links between the two phenomena \citep[e.g.][]{farrah12}. 

The most important issue to establish now is whether a causal relationship exists between star formation and AGN activity, specifically, to determine how the two phenomena regulate each other, and the overall impact of an AGN on its host galaxy. In contrast to observations at a single wavelength band, multi-wavelength studies of the incidence of AGN and star formation in dusty galaxies can provide extensive information on the manner in which the two phenomena coexist, as they are in general sensitive to nuclear activity over a wide range in obscurations \citep[for recent works in the literature see e.g.][]{sajina12,kirkpatrick12,snyder12}. 
Selection in the infrared favours sources with intense AGN or starburst activity, as infrared (IR) spectral energy distributions (SEDs) are likely to be dominated by these two processes: strong mid-infrared (MIR) continua imply AGN activity with hot dust reradiating UV/optical photons \citep[e.g.][]{laurent00,hatzimi05} while strong polycyclic aromatic hydrocarbons (PAHs) in the MIR are signatures of intense star formation \citep[e.g.][]{lutz98,rigopoulou99,farrah08,fadda10}. Silicate absorption at 9.7 \mum is an ambiguous feature, since it can be attributed to both phenomena, although type 1 (i.e., unobscured) AGN are likely to have this feature in emission, or to have featureless continua at this wavelength. And while the relative contributions of the two mechanisms can sometimes be quantified on the basis of broad band photometry alone, individual features can be smeared out, making MIR spectroscopy indispensable for accurate studies.

The combination of the {\it Spitzer} \citep{werner04} and {\it Herschel} \citep{pilbratt10} observatories, especially deep surveys, tracing low and high redshift objects over wide ranges in luminosity, has the potential to make dramatic advances in the study of the AGN-starburst connection.
In this work we explore the impact of the presence of an AGN on the mid- and far-infrared properties of dusty galaxies and how simultaneous AGN and starburst activities affects these properties. We do so using a sample extracted from the HerMES ({\it Herschel} Multi-tiered Extragalactic Survey\footnote{http://hermes.sussex.ac.uk}; \citealt{oliver12}) population, using HerMES SPIRE \citep{griffin10} data as well as a wealth of  publicly available ancillary data, including {\it Spitzer}/IRS spectra \citep{houck04}, of hundreds of extragalactic sources lying in the HerMES fields. 
The paper is structured as follows: Sec. \ref{sec:sample} describes the sample of HerMES objects with IRS spectra and the populations it represents, with measurements applied to the IRS spectra shown in Sec. \ref{sec:specmeas}. Sec. \ref{sec:sedfit} describes the multi-component SED fitting method used, detailing the novelties introduced with respect to previous versions of the code as well as describing  the impact that inclusion of the IRS spectra spectra has on the SED fitting of AGN. Sec. \ref{sec:results} discusses our principal results on the AGN and starburst components in the MIR, the various star formation rate (SFR) estimates and the hot and cold dust components. Finally, Sec. \ref{sec:conclusions} summarises our approach and findings. Throughout this work we assume a $\Lambda$CDM cosmology with $\Omega_{\Lambda}=0.7$ and $\Omega_M$=0.3, and a Hubble constant, H$_0$, of 72 km/s/Mpc.

\section{HerMES sources with IRS spectra}\label{sec:sample}

For the purpose of this work, we select a sample consisting of 375 sources, each detected at $>$ 3$\sigma$ at 250 \mum, and with an IRS spectrum and  spectroscopic redshift measurement (either from an optical or the IRS spectrum) available. As explained in \cite{roseboom10}, the $\sigma$ value considered here includes both the instrumental noise, $\sigma_{\rm inst}$, which is between $\sim$8 and 15.5 mJy depending on the fields \citep[for more details see Table 5 of][]{oliver12} at 3$\sigma_{\rm inst}$ noise level, and the confusion noise, $\sigma_{\rm conf}$ equal to 3.8 mJy at 250 \mum with a 3$\sigma_{\rm conf}$ cut \citep[for more details see][]{nguyen10}. All sources lie in the four northern HerMES fields, namely Bootes HerMES, FLS, Lockman Swire (LS) and ELAIS N1 SWIRE (EN1), covered by the first data release\footnote{http://hedam.oamp.fr/HerMES/release.php} (for a detailed description of the fields and their coverage see \citealt{oliver12}). 
HerMES is a legacy survey conceived to cover about 380 deg$^2$ of nested fields in the most commonly observed extragalactic areas in the sky: fields were chosen among those with the best ancillary data. We therefore have at least 6 photometric points for each source with which to build the SEDs  from optical to the far-infrared (FIR) wavelengths. 

The IRS spectra used in this work are taken from the Cornell AtlaS of {\it Spitzer}/Infrared Spectrograph project (CASSIS\footnote{http://cassis.astro.cornell.edu/atlas/}; \citealt{lebouteiller11}),  which recently made available the reduced low resolution spectra (in two low-resolution modules with a resolving power of R $\sim$ 60-120) of about 11000 sources ever observed with the {\it Spitzer} InfraRed Spectrograph (IRS). 
Henceforth, we refer to this sample as the HerMES/IRS sample. The entire sample has IRAC 3.6 and 4.5 \mums, and MIPS 24 \mum counterparts: $\sim$90 per cent of the objects have been detected at 5.8 and 8.0 \mums, and 77 and 43 per cent of them were also detected at the MIPS 70 and 160 \mums, respectively. SPIRE (The Spectral and Photometric Imaging Receiver, containing a imaging photometer operating at 250, 350 and 500 \mum on board of  the {\itshape Herschel} Space Observatory) fluxes are estimated from scan maps via linear inversion methods, using the positions of known 24 \mum sources as priors \citep[see][]{roseboom10}. As mentioned above, sources are included in the HerMES/IRS sample only if the 250 \mum flux value, S$_{250}$, is greater than 3$\sigma$. No other cut is imposed based on the SPIRE fluxes: 350 and 500 \mum non-zero fluxes are available for 98 and 84 per cent of the HerMES/IRS sample (72 and 35 per cent above 3$\sigma$, respectively). Finally, {\itshape ugriz} photometry from the Sloan Digital Sky Survey Data Release 7 (SDSS DR7; \citealt{abazajian09}) is available for 73 per cent of the sample, with the remaining objects being undetected by SDSS.

\begin{figure}
\centerline{
\includegraphics[width=7cm,angle=270]{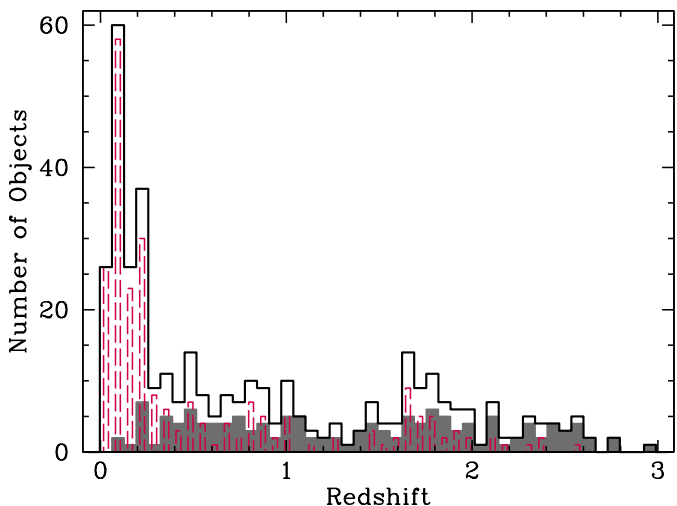}}
\caption{Redshift distribution of the sample. The red dashed and grey shaded histograms correspond to the starburst- and AGN-dominated subsamples: the definition of these subsamples will be described in detail in Sec. \ref{sec:specmeas}.}
\label{fig:zhisto}
\end{figure}

Fig. \ref{fig:zhisto} shows the redshift distribution of the HerMES/IRS sample in black, spanning the range from z$\sim$ 0.014 to 2.99. The various peaks in the distribution reflect the different selection criteria of each of the subsamples present in CASSIS; dashed and shaded regions correspond to the starburst- and AGN-dominated subsamples and will be discussed in the next section. 

\begin{figure}
\centerline{
\includegraphics[angle=-90,width=9cm]{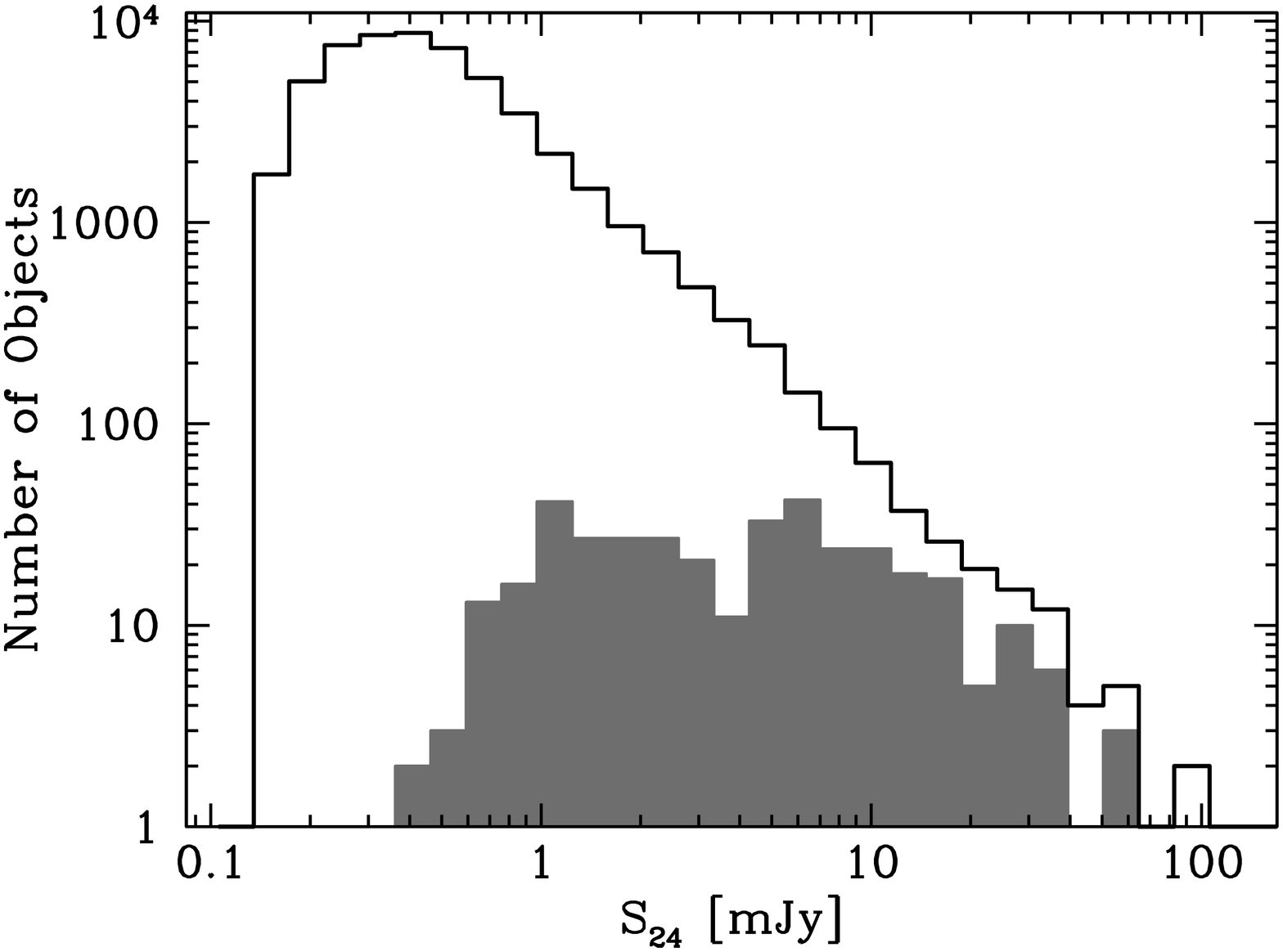}}
\centerline{
\includegraphics[angle=-90,width=9cm]{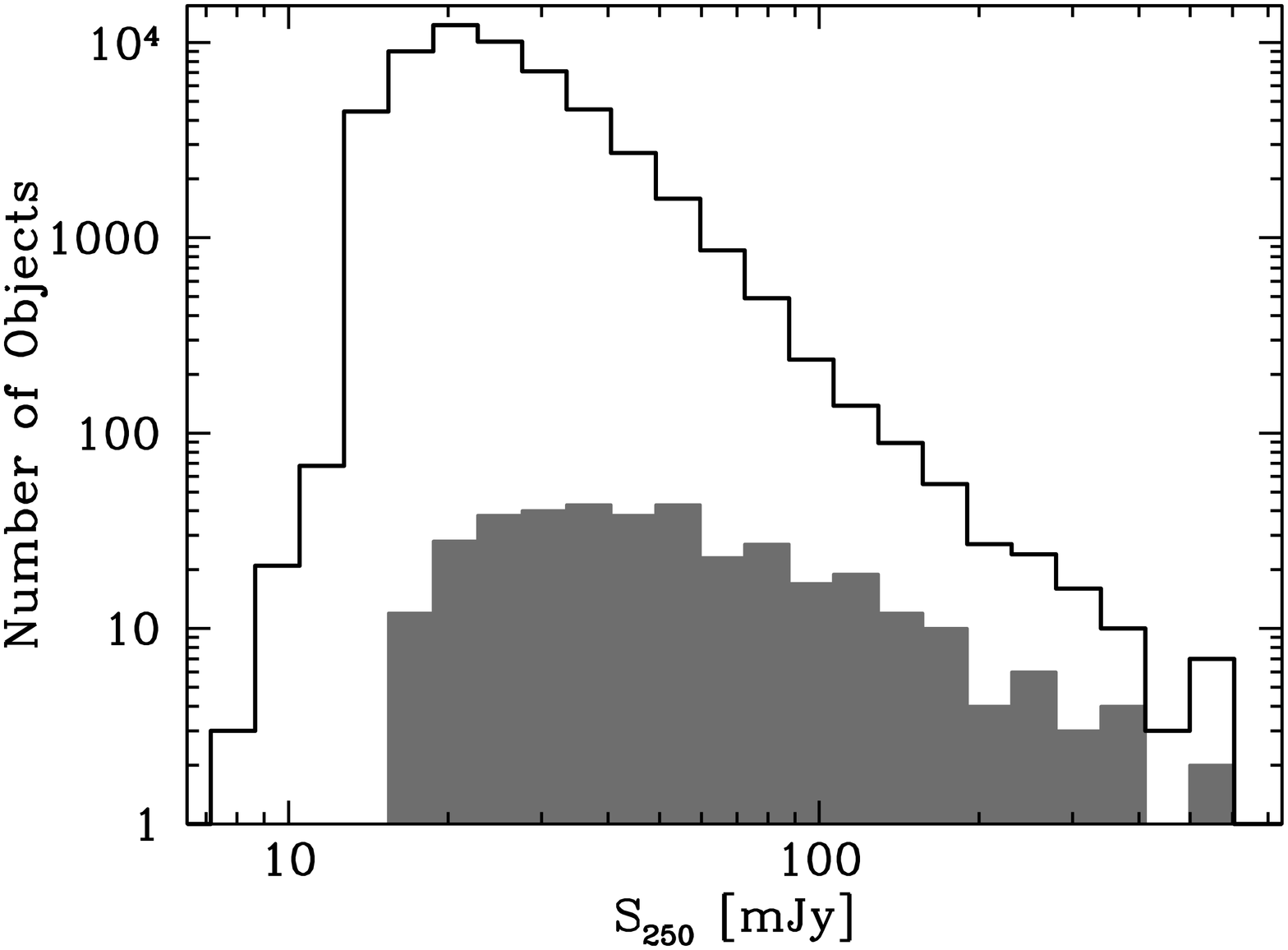}}
\centerline{
\includegraphics[angle=-90,width=9cm]{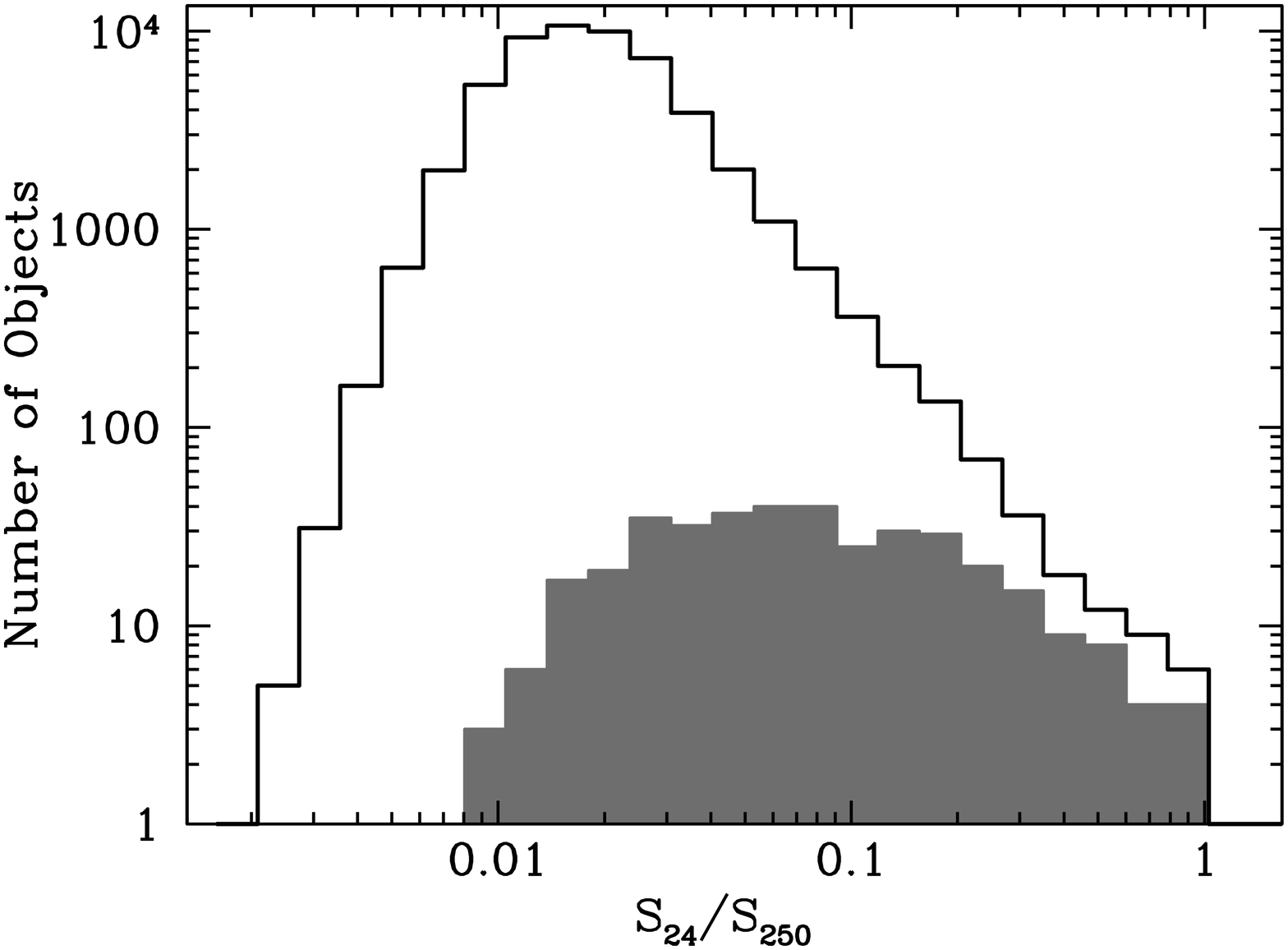}}
\caption{S$_{24}$ \mum (top), S$_{250}$ \mum (middle) and S$_{24}$/S$_{250}$ (bottom) distributions of the full HerMES population in the four fields (solid line) and those of the HerMES/IRS sample (shaded region).}
\label{fig:fluxhistos}
\end{figure}

The IRS spectra come from a compilation of various small and larger {\it Spitzer} programmes with objects including dusty galaxies, LIRGs and ULIRGs, normal star forming galaxies and AGN. The details of programs contributing with more than 10 objects to the final sample are summarized in Table \ref{tab:spitzerprog}. The majority of sources ($\sim 56$ per cent of the entire sample) are selected at 24 \mums, with a flux cut at 24 \mums, S$_{\rm 24}$, depending on the details of the individual programmes: S$_{\rm 24} > 0.5, 0.7, 0.9, 1.0$ mJy or higher for the $\sim$ 15, 16, 7, 4 and 4 per cent of the sources of the total sample, respectively. Another $\sim13$ per cent of the entire sample is selected at 70 \mums. Note that most of the IRS samples are flux limited. A notable exception is the sample of \cite{yan07}, where both 24 \mum flux and a 8-24 \mum colour cuts are applied. This selection biases the sample in favour of AGN type and MIR bright sources, avoiding also objects with strong silicate absorption features. Due to its selection criteria the sample of \cite{weedman09} is not complete, with sources classified as starburst by optical spectra not being detected by IRS. The sample of \cite{farrah08} is biased towards ULIRGs with ongoing star formation to the detriment of those containing luminous AGN. Instead, the sample of \cite{houck07} represents a complete distribution of IR extragalactic sources including those with featureless spectra and those with strong PAH and/or silicate features. The flux-selected samples at 70 \mums, as that of \cite{farrah09}, are less affected by biases with the selection being sensitive both to starburst and AGN sources. Even though the presence of possible different small selection biases in the different subsamples, the HerMES/IRS sample is a large sample, spanning several orders of magnitude in luminosity, and the properties of its sources match the entire range of the IR bright HerMES population. 
Fig. \ref{fig:fluxhistos} shows the S$_{24}$ \mum (top), S$_{250}$ \mum (middle) and S$_{24}/S_{250}$ (bottom) distributions of the full HerMES population in the four fields (solid line) and as well as those of the HerMES/IRS sample (shaded region).

%
%
%
%
%
%
%
%
%
%

\begin{table*}
\small
\begin{center}
\begin{tabular}{cccccc}
\hline
ID & PI  & Selection & Reference & Field & Number of Sources \\
\hline
15 & Houck  & S$_{24}>$0.7 mJy & & LS, Bootes, FLS, EN1 & 47\\
16 & Houck  & S$_{24}>$0.7 mJy &  & Bootes & 14\\
3748 & Yan & S$_{24}>$1 mJy &  & FLS & 15\\
20113 & Dole & 70\mums-selected & & Bootes & 15\\
20128 & Lagache & 24\mums-selected &  e.g. \cite{weedman09} & Bootes, FLS & 21\\
20629 & Yan & S$_{24}>$0.9 mJy & \cite{yan07} & FLS & 28 \\
30364 & Houck & S$_{24}>$ 0.5 mJy & \cite{farrah08} & LS & 25\\
40038 & Houck  & S$_{24}>$ 6 \& mJy S$_{24}>$ 10 mJy & \cite{houck07} & Bootes, FLS & 14\\
40539 &  Helou & 0.5 mJy $<$S$_{24}<$100 mJy & e.g. \cite{weedman09} & LS, FLS, EN1 & 30\\
50666 & Farrah & S$_{70}>$20 mJy & \cite{farrah09} & LS & 32\\
\hline
\end{tabular}
\end{center}
\caption{Major contributors to the HerMES/IRS sample. Shown are the {\it Spitzer} programme ID, the PI of the proposal, the primary selection criteria and the reference, whenever available, the corresponding fields and number of sources.} 
\label{tab:spitzerprog}
\end{table*}

\section{MIR AGN- and starburst-dominated objects}\label{sec:specmeas}


PAH features are commonly used to quantify AGN and star formation activity,  as well as the relative contribution of the two physical processes to the energy output of dusty galaxies in the MIR \citep{lutz96,genzel98,rigopoulou99,armus06,veilleux09}. Indeed, the luminosity of PAH, L$_{\rm PAH}$, allows for an estimate of the star formation rate, SFR$_{\rm PAH}$ \citep[][see Sec. \ref{sec:results} for more details]{brandl06,houck07,pope08,hernan09}. Previous work has established that the EW$_{\rm PAH}$ of the 6.2 and 11.3 \mum bands correlate with the relative contributions of the AGN and starburst to the bolometric output of the galaxy \citep[e.g.][]{laurent00,spoon07}. \cite{hernan11} estimated that EW$_{\rm PAH}$=0.2 \mum represents roughly equal contributions from the AGN and starburst to the bolometric luminosity (valid for either of the two PAH features), with the AGN dominating at lower values and the starburst dominating at higher values. For the purpose of this work, we use the 11.3 \mum PAH feature, and use that 6.2 \mum only in the absence of the former (necessary in $\sim$ 10 per cent of the objects).

We measured the equivalent widths of the PAH features, EW$_{\rm PAH}$, at 6.2 and 11.3 \mums, as well as their luminosities from the IRS spectra using the procedure described in \cite{hernan11}. Briefly, we select continuum bands 0.2 \mum wide at both sides of each PAH feature (centered at 5.9 and 6.6 \mum for the 6.2 \mum PAH feature; 10.88 and 11.78 \mum for the 11.3 \mum one) and interpolate linearly between them to estimate the continuum under the feature. We subtract the linear continuum and integrate the residual in a band of width 0.5 \mum centered at the nominal wavelength of the feature to obtain the flux in the PAH band. As explained in \cite{hernan11}, our selection of a narrow integration band and nearby continuum bands help to maximize the signal-to-noise (S/N) and to reduce the uncertainty in the underlying continuum. Flux lost in the wings of the PAH bands and contamination to the continuum are corrected for by assuming the PAH feature has a Lorentzian profile with a full width at half-maximum (FWHM) of 0.2 \mums. Simulations indicate that a 10 per cent increase in the FWHM causes a 5 or 6 per cent drop in the PAH flux, with no observable dependency on the EW$_{\rm PAH}$ or S/N of the spectrum. Uncertainties in the PAH flux and the underlying continuum for each source are estimated by performing Monte Carlo simulations. 

Extinction has complex interactions with the PAH features: in highly obscured sources, the 6.2 \mum band is often attenuated or even entirely suppressed due to a water ice absorption band at 6 \mum \citep[e.g.][]{spoon04,imanishi07,sajina09}. The 11.3 \mum band is embedded inside the much wider silicate feature at 9.7 \mum (see Sec. \ref{sec:ssil}), and the effect of extinction on it depends on whether the obscuration producing the silicate feature affects only the AGN or also the starburst component. In the first case, continuum emission from the AGN gets diminished, and the EW$_{\rm PAH}$ of the feature at 11.3 \mum is accordingly boosted. In the second, both the continuum and the feature are equally suppresed, and therefore the EW$_{\rm PAH}$ remains unchanged. Since most of the sources in the sample have moderate to low values of the strength of the silicate feature (see Fig. \ref{fig:ssilspec}), our results are largely unaffected by extinction.

\begin{figure}
\centerline{
\includegraphics[width=7cm,angle=270]{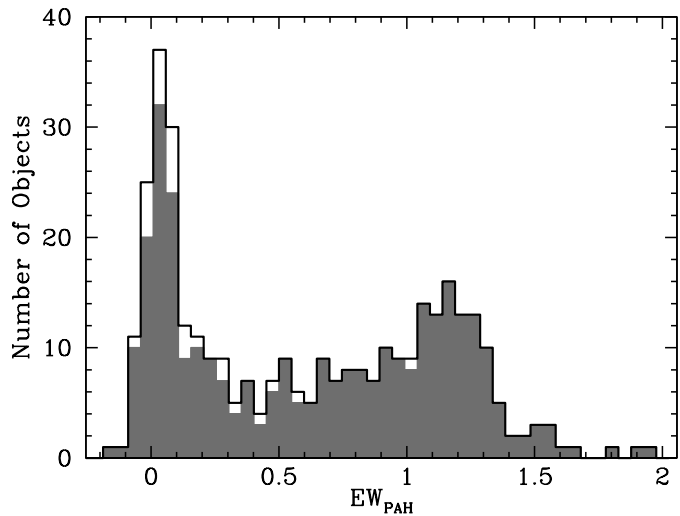}}
\caption{Distribution of EW$_{\rm PAH}$ (that at 11.3 \mum shown in grey; PAH at 6.2 \mum used when the 11.3 \mum data was not available), measured from the IRS spectra.}
\label{fig:ewpah}
\end{figure}

The distribution of EW$_{\rm PAH}$ is shown in Fig. \ref{fig:ewpah}, where the grey histogram corresponds to the feature at 11.3 \mums. In the HerMES/IRS sample we found 45\% and 55\% MIR AGN- and starburst- dominated objects, i.e. with EW$_{\rm PAH} < 0.2$ and $>0.2$, respectively. Hereafter, when reporting to AGN- dominated (EW$_{\rm PAH} <$ 0.2) or starburst-dominated (EW$_{\rm PAH} >$  0.2) objects we refer to the distinction made based on the calculated values of EW$_{\rm PAH}$.

Returning to the redshift distribution of the HerMES/IRS sample (Fig. \ref{fig:zhisto}), we note that very low redshift sources may have angular sizes larger than the width of the slits used for IRS spectroscopy (3.6" and 10.5" wide for the short- and long-wavelength modules, respectively). As a consequence, a fraction of the emission from the outer regions of these galaxies may be missed by the IRS observations. This, in principle, could result in an overestimation of the AGN contribution to the galaxy's measured emission, which would translate to an excess of AGN-dominated objects at very low redshifts. As seen in the figure, the more nearby objects ($z\le0.3$) are almost exclusively starburst-dominated (red dashed histogram) with only 10 objects being AGN-dominated.  
The effects of slit width can be assessed by comparing the MIPS 24 \mum flux with that evaluated at 24 \mum from the IRS spectra. We find consistent values for all low-redshift sources in the HerMES/IRS sample, and conclude that aperture effects do not significantly affect our results and conclusions. 
We note that almost all HerMES/IRS galaxies at $z>2$ (Fig. \ref{fig:zhisto}) are AGN-dominated (shaded histogram): this is not a physical result but is due to the selection of the IRS targets that constitute the various subsamples of our HerMES/IRS sample.

\begin{figure}
\centerline{
\includegraphics[width=9.5cm]{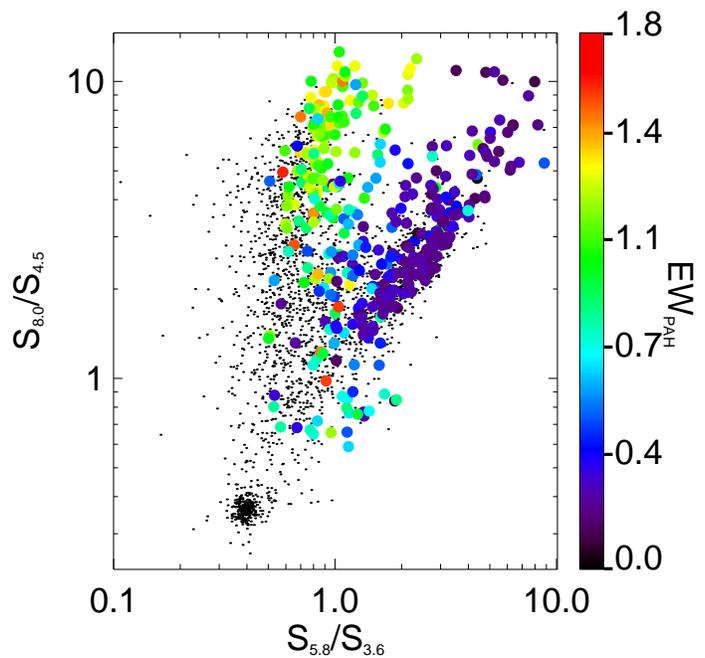}}
\caption{IRAC colours of the HerMES/IRS sources. The points are colour-coded as a function of  EW$_{\rm PAH}$. Black points show the density of objects in the FLS field, with stars clustering in the lower left corner of the plot.}
\label{fig:iraccol}
\end{figure}

It is well known that AGN-dominated sources tend to concentrate in a particular region of the IRAC colour-colour diagram \citep[e.g.][]{lacy04,lacy07,stern05,donley12}. In fact, Fig. \ref{fig:iraccol} shows the (observed) IRAC colours of the HerMES/IRS sources, colour-coded based on the value of EW$_{\rm PAH}$, justifying the use of the latter to separate between AGN- and starburst-dominated objects. \cite{hatzimi09} showed that the position of AGN on this colour-colour diagram depends on the relative contribution of their starburst content to the MIR, going from the diagonal locus (defined by the MIR AGN continuum) for pure AGN to the vertical S$_{5.8}$/S$_{3.6}$ locus, as the starburst contribution increases. They also showed that the position of AGN is almost independent of redshift up to a redshift of $\sim 3$ (see their Fig. 2), a value corresponding to the most distant object in our HerMES/IRS sample. AGN-dominated objects in the HerMES/IRS sample, as classified by the value of EW$_{\rm PAH}$, show the same behavior and lie on or around the power-law slope; as the starburst component becomes increasingly dominant, the position of sources moves towards the vertical S$_{5.8}$/S$_{3.6}$ locus. The black points denote objects in one of the selected fields (FLS), with stars clustering in the lower left corner of the plot \citep[see e.g.][]{sajina05,hatzimi09}. 

\section{SED fitting}\label{sec:sedfit}
\begin{figure}
\centerline{
\includegraphics[angle=270,width=8cm]{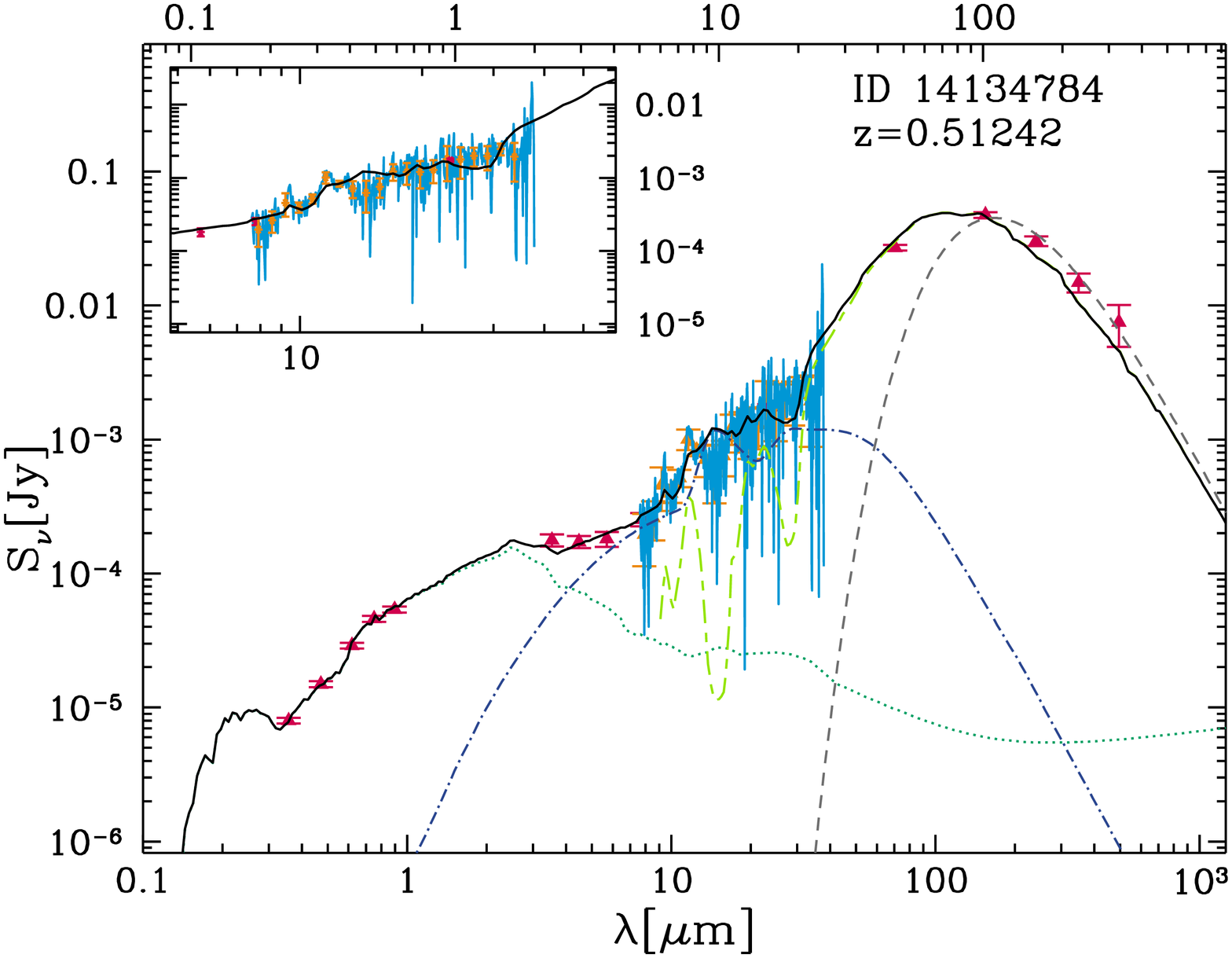}}
\centerline{
\includegraphics[angle=270,width=8cm]{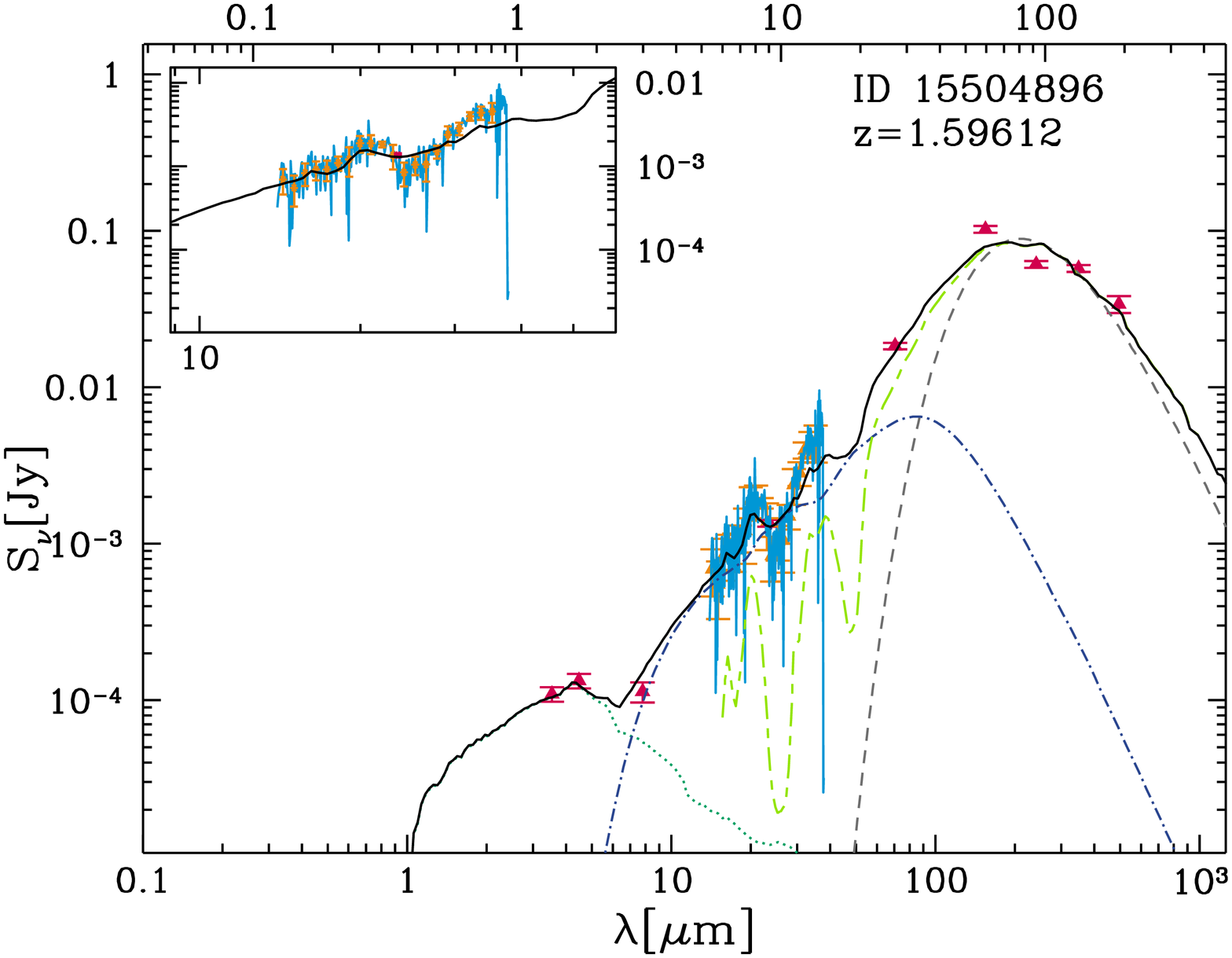}}
\caption{Example of a best-fit to a low (top) and a high (bottom) redshift object ({\itshape Spitzer} ID 14134784 and 11867904, respectively). The IRS spectrum (turquoise; see also the insert figure) and photometric data (red symbols) are reproduced using the three emission components: SSPs (dotted dark green line), AGN torus (dot dashed blue line) and starburst (dashed light green line), giving the total model emission (in black). The dashed grey lines show the best-fit modified black body emission, fitted at a second step. The top x-axis shows the rest-frame wavelength.} 
\label{fig:fits}
\end{figure}

We use a routine described in detail in \cite{hatzimi08,hatzimi09} to fit SEDs to the HerMES/IRS sample data. 
The observed SED of each object is compared to a set of model SEDs by means of a standard $\chi^2$ minimisation. An addition to the original code is the possibility to simultaneously fit photometry {\it and} an IRS spectrum for each object, as described in Sec. \ref{sec:specphotfit}.
As shown in Fig. \ref{fig:fits}, the model SEDs (black line) are the sum of three components: a stellar (dotted dark green), an AGN (dot dashed blue) and a starburst (dashed light green) component. The stellar component is itself the sum of Simple Stellar Population (SSP) models of different age, all having a common (solar) metallicity, and is built up using the Padova evolutionary tracks \citep{bertelli94}, a Salpeter initial mass function (0.15-120 $M_{\odot}$) and the \cite{jacoby84} library of observed stellar spectra in the optical domain. 
The AGN component consists of the emission from the primary source and the emission reprocessed by dust, distributed in a continuous fashion (as opposed to clumps) in a toroidal or flared-disk shaped region around the primary source and described in detail in \cite{fritz06}. In this work, we use the updated AGN torus model grid presented in \cite{feltre12}. Finally, the starburst component is represented by a library of starburst templates that are used to reproduce the detailed PAH features of the IRS spectra. Sources in the library include Arp220, M82, M83, NGC1482, NGC4102, NGC5253 and NGC7714. Due to their empirical nature, the starburst templates cannot be used to compute accurate values of physical quantities such as the mass of cold dust and its temperature. For this reason, and as a second step, we fit the FIR data points  ($\lambda >$ 100\mums) of the SEDs to a modified black body emission, as described in detail in Sec. \ref{sec:bb}. 

The SED fitting procedure described above also takes into account the interstellar dust and the extragalactic flux attenuations. Extinction by dust in the interstellar medium is modelled as a uniform slab in front of the stars. It is parameterised by an E(B-V) value and an assumed galactic extinction curve taken from \citep{cardelli89}. We checked whether adopting other prescriptions for the dust extinction, e.g. the extinction curve as proposed by \cite{calzetti94} and widely used in studies of actively star-forming objects, would alter our results: we are only fitting five broad band optical data points (SDSS) and found that using the Calzetti model neither changed significantly the stellar masses estimations, nor provided a better fit to the data (as quantified by the value of $\chi^2$).  Finally, we model attenuation by the intergalactic medium using the \cite{madau95} law. 

\subsection{Spectrophotometric fitting}\label{sec:specphotfit}

We perform the SED fitting using photometric data points (typically SDSS $urgiz$, {\it Spitzer} IRAC and MIPS, and {\it Herschel} SPIRE fluxes: red points in Fig. \ref{fig:fits}) and the IRS spectra simultaneously, and use standard $\chi^2$ minimization to determine the best fit. Each IRS spectrum (shown in turquoise in Fig. \ref{fig:fits}) is divided in a predefined number of bands (20), with the mean flux and the respective error calculated in each band. These fluxes are then handled in the same way as the photometric data points. The best-fit model comprises the combination of the models providing the lowest values of the reduced $\chi^{2}$, $\chi^{2}_{\nu}$, which is by definition the $\chi^2$ divided by the number of degrees of freedom, in turn the number of data points - the number of model parameters.

An AGN component was required to reproduce the observed SED for about 96 per cent of the HerMES/IRS sources. For the $\sim$25 per cent of these, no stellar component was needed because in these objects the light of the AGN outshines that of the host galaxy in the UV/optical, and the sources are therefore characterised as unobscured AGN. Inconspicuous PAH features (EW$_{\rm PAH} < 0.2$ ) further demonstrate that the AGN dominates the MIR emission of such sources (for more details see Sec. \S \ref{sec:results}).

Important physical quantities can be derived from the SED fitting including the AGN accretion luminosity, L$_{\rm acc}$, i.e. the soft X-ray, UV and optical luminosity coming from the accretion disc, the infrared luminosity, L$_{\rm IR}$, defined as the integrated flux between 8 and 1000 \mums, and the relative contribution of the AGN and the starburst, L$_{\rm SB}$. The latter quantity can be used to estimate the obscured star formation rate, SFR$_{\rm FIR}$, using the calibration derived in \cite{kennicutt98}, in which the bolometric luminosity of stars younger than 100 Myr is assumed to be re-emitted in the IR \citep[see][and Sec. \ref{sec:results} for more details]{kennicutt98,leitherer95}.

\subsubsection{On the effects of the IRS spectra on the SED fitting} \label{sec:ssil}
\begin{figure}
\centerline{
\includegraphics[width=7cm,angle=270]{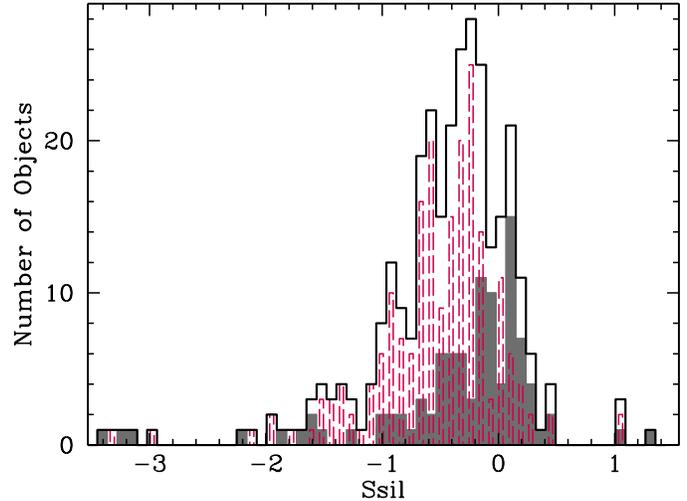}}
\caption{ 
Distribution of the strength of the silicate feature at $\sim$9.7 \mums. The grey shaded and red dashed histograms illustrate AGN- and starburst-dominated sub-samples, defined as such by EW$<0.2$ and EW$\ge0.2$, respectively.}
\label{fig:ssilspec}
\end{figure}

The strength of the silicate feature at 9.7 \mums, $S_{sil}$ is defined by \cite{pier92} as:
\begin{equation}
S_{sil} = ln[F(\lambda_p)/F_C(\lambda_p)]
\label{eq:ssil}
\end{equation}
Where $F(\lambda_p)$ and $F_C(\lambda_p)$ are the feature's peak flux density and the underlying continuum respectively. It follows from this that a negative value of $S_{sil}$ indicates that the silicate feature is in absorption.
The distribution of $S_{sil}$ for the sample is shown in Fig. \ref{fig:ssilspec}.
A large fraction (75 per cent) of the objects with the feature in emission are AGN-dominated based on calculated values of EW$_{\rm PAH}$.

\begin{figure}
\centerline{
\includegraphics[width=7cm,height=7cm]{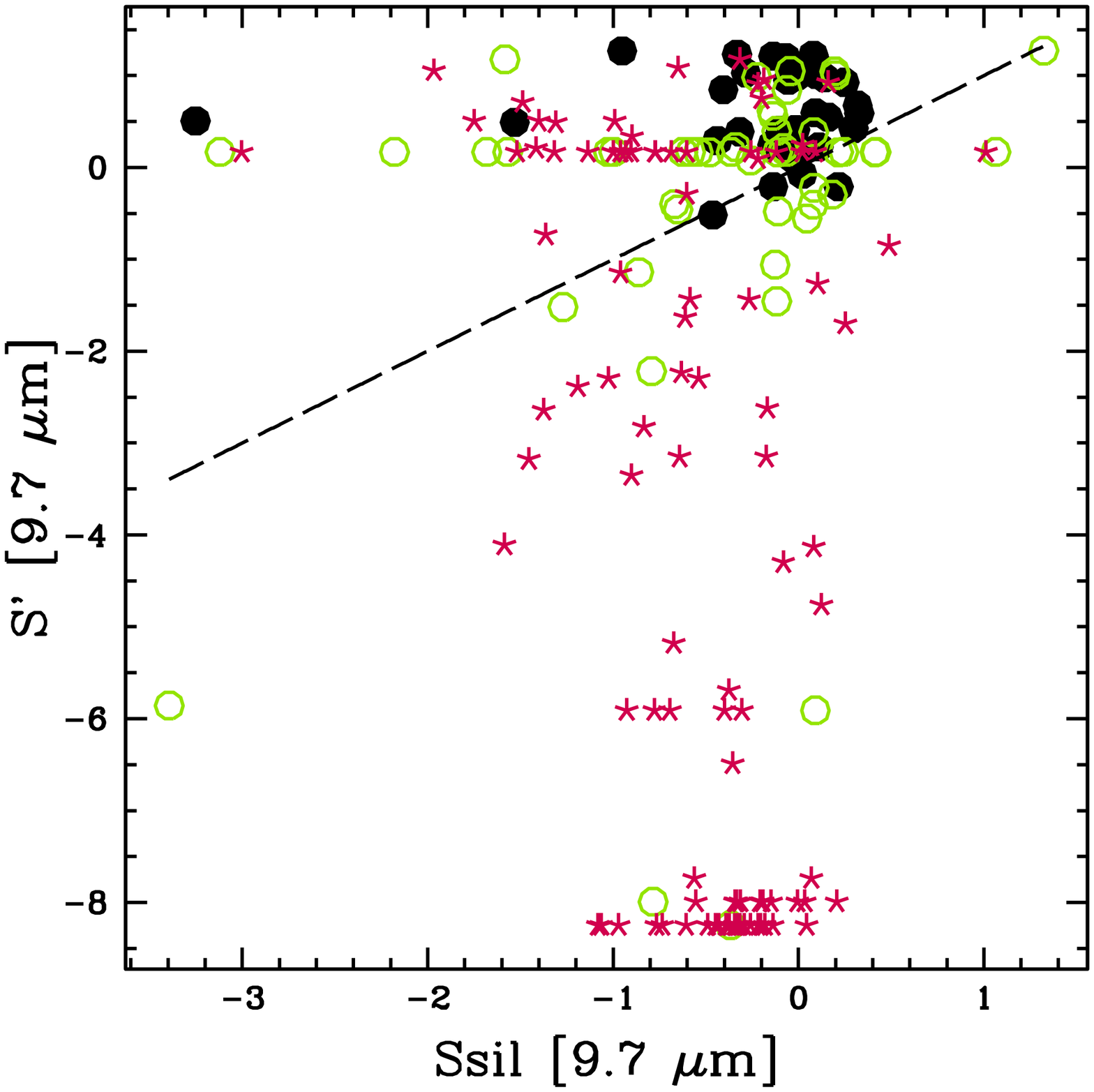}}
\centerline{
\includegraphics[width=7cm,height=7cm]{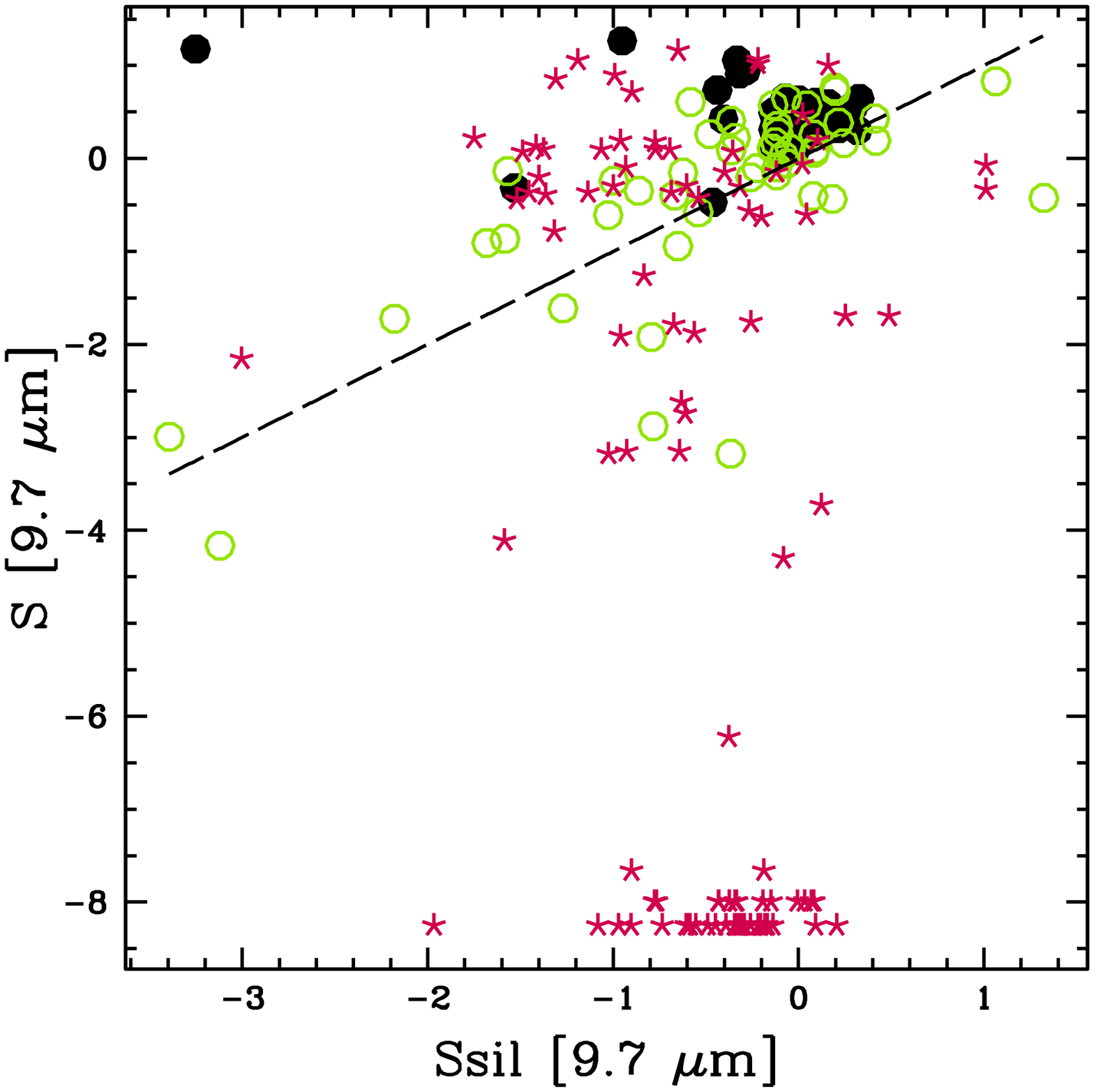}}
\centerline{
\includegraphics[width=7cm,height=7cm]{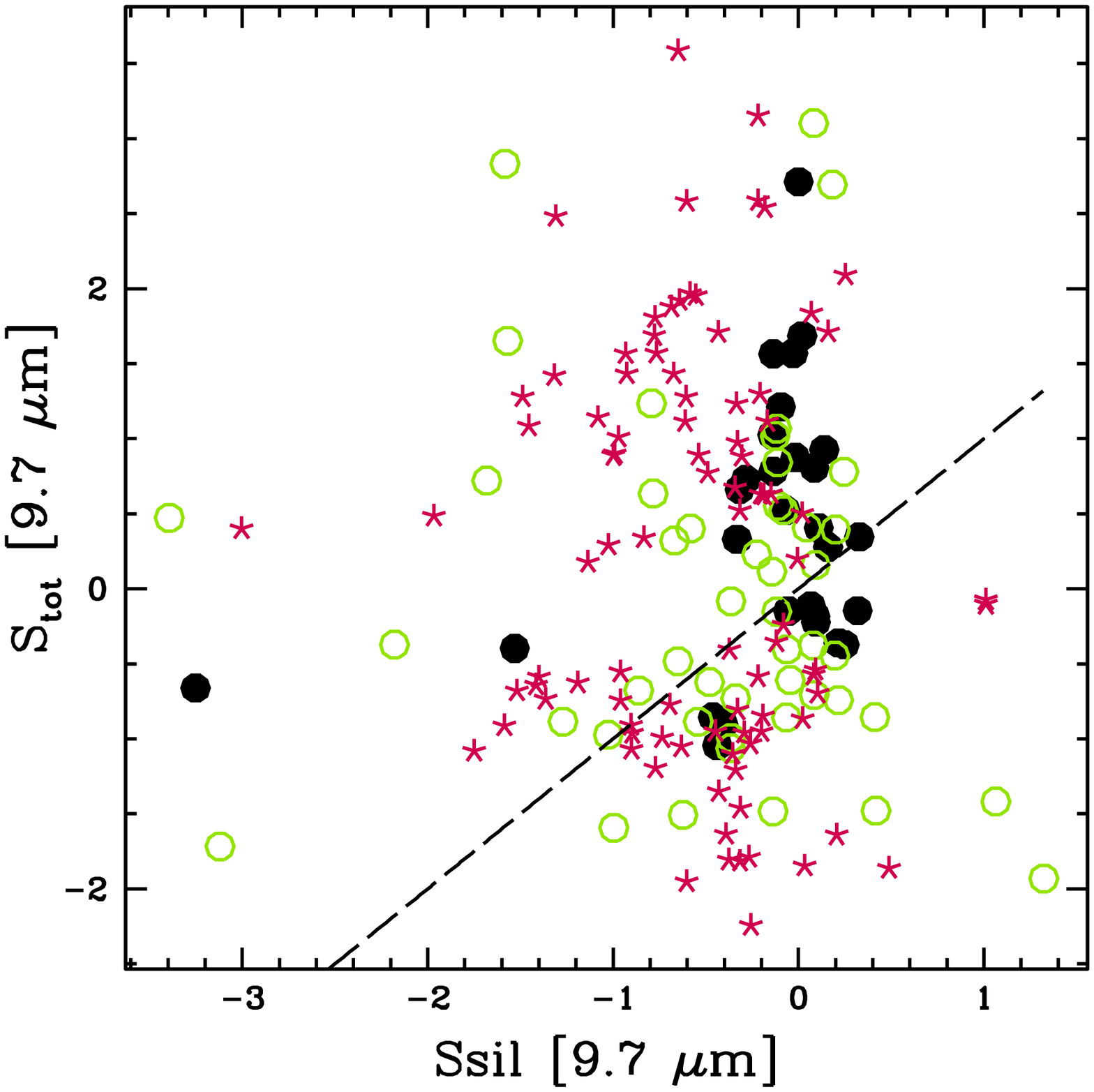}}
\caption{The strength of the silicate feature at 9.7 \mum measured from the best-fit model versus that measured from the IRS spectra. In the top panel, $S'$ was measured on the best-fit torus models obtained by fitting photometry alone, while $S$ (middle panel)  was measured from the best-fit torus model when fitting simultaneously photometric data and IRS spectra. In the bottom panel, $S_{tot}$ is the measured strength of the silicate feature on the total model, i.e. the sum of the AGN and the starburst component. The filled black (open green) circles correspond to AGN-dominated objects without (with) a stellar component (SSPs), while red stars correspond to starburst-dominated objects.} 
\label{fig:sil}
\end{figure}

$S_{sil}$ is a property of the hot dust that relates to its opacity and geometry and cannot be constrained with photometric data points alone. In the following we determine the impact of including the IRS spectra in the evaluation of $S_{sil}$. 
We measure this quantity using the best-fit torus model for all the objects, 
with and without including the IRS spectra in the SED fitting ($S$ and $S'$, respectively). We then compare the values, $S$ and $S'$ with the values of $S_{sil}$ measured directly from the IRS spectra, as shown in Fig. \ref{fig:sil}, top and middles panels, respectively. In the presence of both AGN and starburst components, the SED fitting has to reproduce the silicate feature by adding the two components, hence measuring $S$ on the best fit torus model alone might be irrelevant, especially for objects with strong starburst contribution. Indeed, in this case, $S$ has to be measured on the total (i.e., AGN+starburst) model ($S_{tot}$, shown as a function of $S_{sil}$ in the bottom panel of Fig. \ref{fig:sil}). The circles correspond to all the AGN-dominated objects (EW$_{\rm PAH} < 0.2$), with the open (filled) symbols representing objects with (without) a stellar component. Finally, the stars indicate starburst-dominated objects (EW$_{\rm PAH}>0.2$).


Fig. \ref{fig:sil} shows several interesting consequences of including IRS spectra in SED fitting. A simple comparison between the top and middle panels shows that the cluster of points at $S'\sim0.0$ for objects with a stellar component and $S_{sil}<0.0$ to vanish. For these objects including IRS spectra in the fit returns torus models with a silicate feature whose strength matches more closely that measured in the IRS spectra. Since $S$ depends strongly on the model parameters, a more accurate measurement, as in the presence of the IRS spectra, implies better constraints on the model parameters. 
Furthermore, the clustering of unobscured AGN around the 1:1 line becomes tighter, although SED fitting still tends to favour torus models with silicate emission that is stronger than observed. 
In starburst-dominated objects the IRS spectra do not improve the constraints on the silicate feature of the torus models, as can be seen in the top and middle panels of Fig. \ref{fig:sil}, where both $S$ and $S'$ present very deep absorption ($\sim$ -8.0). Indeed, as the starburst component starts becoming important, the points disperse even in the presence of IRS spectrum and $S_{tot}$ has to be considered instead (bottom panel of Fig. \ref{fig:sil}). Despite a non-negligible scatter, likely due to the noise of the spectra, model and spectral measurements now cluster around the 1:1 line, particularly tightly for unobscured AGN (filled circles).
Our findings show that including the IRS spectra in the SED fitting helps to better constrain the AGN component for objects whose optical/MIR SEDs are dominated by the AGN; as a consequence, better constraints on the starburst component are obtained as well.

\subsection{Fitting a modified black body to the FIR points}\label{sec:bb}

The starburst templates used to fit the observed emission, due to their empirical nature, cannot be used to derive accurate estimates of the physical properties of the cold dust component, such as mass and temperature. For this reason data at $\lambda > 100$ \mum are fitted separately, assuming that dust is emitting as a single-temperature, modified black-body. Dust emissivity is modelled as power law $k_{\nu}=k_0\nu^\beta$, with $k_0$ a normalization factor such that $k_{350\mu m}=0.192$ m$^2$kg$^{-1}$ \citep[][]{draine03}. The modified black body is then expressed in the analytic form:
\begin{equation}
S_{\nu}(\beta,T)=\frac{M_c k_0}{d^2}\left(\frac{\nu}{\nu_{0}}\right)^{\beta}B_{\nu}\left(T\right)
\label{eq:mbb}
\end{equation}
where $M_c$ is the mass of the (cold) dust and $d$ is the luminosity distance. The only two free parameters in the fitting procedure are the dust temperature and mass: we chose to fix the value of $\beta$ at 2, which is consistent with the value commonly used in dust models \citep[][see also \citealt{davies12} for observational evidence]{draine84,li01}. 

The best fit temperatures are found by means of a gradient search method following \cite{fritz12}. A well-sampled SED with photometry that spans the IR peak is required to properly constrain the properties of the cold dust \citep{kirkpatrick12}: fitting using the modified black body has therefore only been applied to objects with at least two SPIRE band detections as well as MIPS 160 \mums, jointly available for $\sim$30 per cent of the HerMES/IRS sample.

\section{AGN and star formation in the MIR and FIR}\label{sec:results}

\begin{figure}
\centerline{
\includegraphics[width=6.5cm,angle=270]{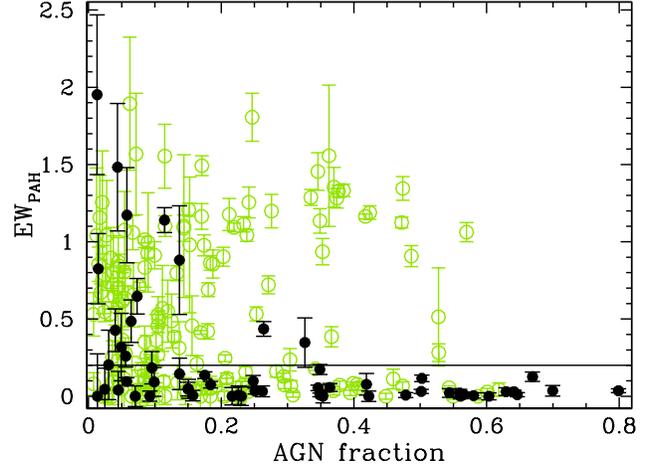}}
\caption{EW$_{\rm PAH}$ as a function of the fractional contribution of the AGN component to the total IR luminosity for all objects to which the SED fitting assigned an AGN component. The filled (open) symbols are objects without (with) an optical stellar component. The black line (EW$_{\rm PAH}$ = 0.2) delimits the MIR AGN- and starburst dominated objects, with the first taking lower values.}
\label{fig:ewagnfrac}
\end{figure}

We use EW$_{\rm PAH}$ to distinguish between AGN- and starburst-dominated objects in the MIR, with the AGN-dominated objects being characterized by values lower than the threshold value, EW$_{\rm PAH}=0.2$. An important question is how this measure compares to the absolute AGN and starburst contributions to the overall emission from the galaxies. Fig. \ref{fig:ewagnfrac} shows EW$_{\rm PAH}$ measured from the IRS spectra as a function of the AGN fractional contribution to L$_{\rm IR}$, in turn derived from the SED fitting (as already mentioned, an AGN component was necessary to reproduce the SEDs for 85 per cent of the HerMES/IRS sample). The green open circles denote the AGN with a stellar component while the black filled circles  AGN without one (see Sec. \ref{sec:specphotfit}). The two quantities are not expected to be correlated, as the contribution of the AGN component to the MIR is not necessarily representative of its contribution to the total energy output, there is clearly an avoidance zone, however: objects with a high fractional AGN contribution to the L$_{\rm IR}$ do not exhibit any PAH features (EW$_{\rm PAH}\sim0$). Furthermore, $\sim$75 per cent of the unobscured AGN (filled circles) also have EW$_{\rm PAH}<0.2$. At the same time, 40 per cent of the AGN with a stellar optical component (open circles) also have EW$_{\rm PAH}<0.2$. Last but not least, some sources present large EW$_{\rm PAH}$ (with values between 1 and 1.5) and, at the same time, a relatively high AGN contribution (between 0.4 and 0.6). These objects are characterized by values of $S_{sil}$ that are around zero, or slightly negative due to absorption suppressing the continuum around the PAH features, and by best fit torus models with large size and high optical depth giving rise to a significative AGN contribution even at longer wavelength. The presence of a starburst emission component is necessary to account for the total far-IR emission. Indeed, it is worth highlighting the presence of a cluster of objects ($\sim 37$ per cent) with very low EW$_{\rm PAH}$ (AGN-dominated in the MIR) and very low AGN fraction, meaning that the total IR luminosity is dominated by the starburst mechanisms. \cite{sajina12} find a class of composite objects ($\sim$ 47 per cent of their sample of 191 24-\mum bright sources in FLS field) showing similar properties, that is high starburst contribution to the L$_{\rm IR} $ along with a very low value of the EW of the PAH feature at 7.7 \mums.  Moreover, we found very few objects ($\sim$8 per cent) having a fractional contribution of the AGN to L$_{\rm IR}$ of more than 50 per cent, confirming once again that the source of the bulk of the IR emission, even in AGN-dominated systems, is star formation. 
The 53 per cent of the HerMES/IRS sample objects presenting a fractional contribution of the AGN to L$_{\rm IR}$ of less then 50 per cent also presents EW$_{\rm PAH}>0.2$ (all but three MIR starburst-dominated sources), i.e. starburst emission dominates both the MIR and the FIR. Our findings are in agreement with those of \cite{kirkpatrick12} where they analyze a sample of 151 24 \mum selected galaxies in the GOODS-N and ECDFS fields. They found most of the starburst-dominated sources in the MIR have a negligible AGN contribution to the FIR, while MIR AGN-dominated objects show various levels of contribution by star formation activity.

\subsection{AGN and the star formation rate from MIR and FIR}

The combination of spectral measurements and SED fitting allows us to estimate the {\it obscured} SFR in the objects of our sample in two different ways, as already described in Secs. \S \ref{sec:specmeas} and \ref{sec:specphotfit}:
\begin{itemize}
\item[]{i)} SFR$_{\rm FIR}$, the obscured star formation rate, is calculated using \cite{kennicutt98}, converting L$_{\rm SB}$ into a value for SFR using SFR$_{\rm FIR}=4.5 \times 10^{-44} \times$ L$_{\rm SB}$, with  L$_{\rm SB}$ in erg/sec. 
\item[]{ii)} SFR$_{\rm PAH}$, derived from the luminosity of the PAH features as measured from the IRS spectra (see \S \ref{sec:specmeas}), and defined as SFR$_{\rm PAH}$ = $1.4 \times 10^{-8} \times$ L$_{\rm PAH[6.2]}$, SFR$_{\rm PAH}$ = $1.52 \times 10^{-8} \times$ L$_{\rm PAH[11.3]}$, where PAH luminosities are expressed in erg/sec  \citep{hernan09}.  We verified that the infrared luminosity derived from the luminosities of the PAH features at 6.2 \mum and 11.3 \mum are comparable, and adopted the value derived for the PAH feature at 11.3 \mum as it is less affected by extinction. We use the PAH luminosity at 6.2 \mum only for objects for which a measurement at 11.3 \mum is not available (about 10 per cent of the objects, as already mentioned in Sec. \ref{sec:specmeas}. 
\end{itemize}

\begin{figure}
\centerline{
\includegraphics[width=7cm]{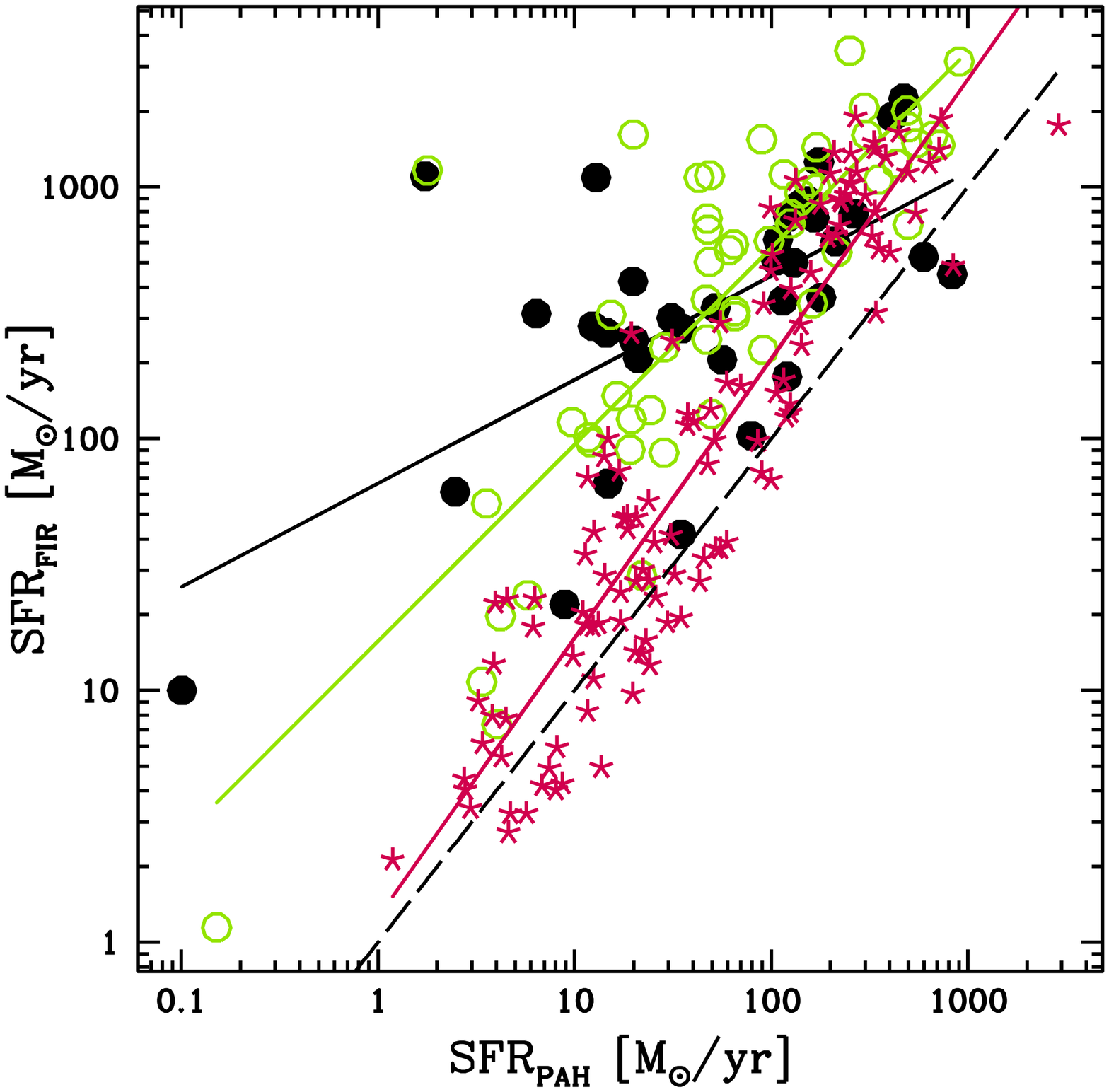}}
\centerline{
\includegraphics[width=7cm]{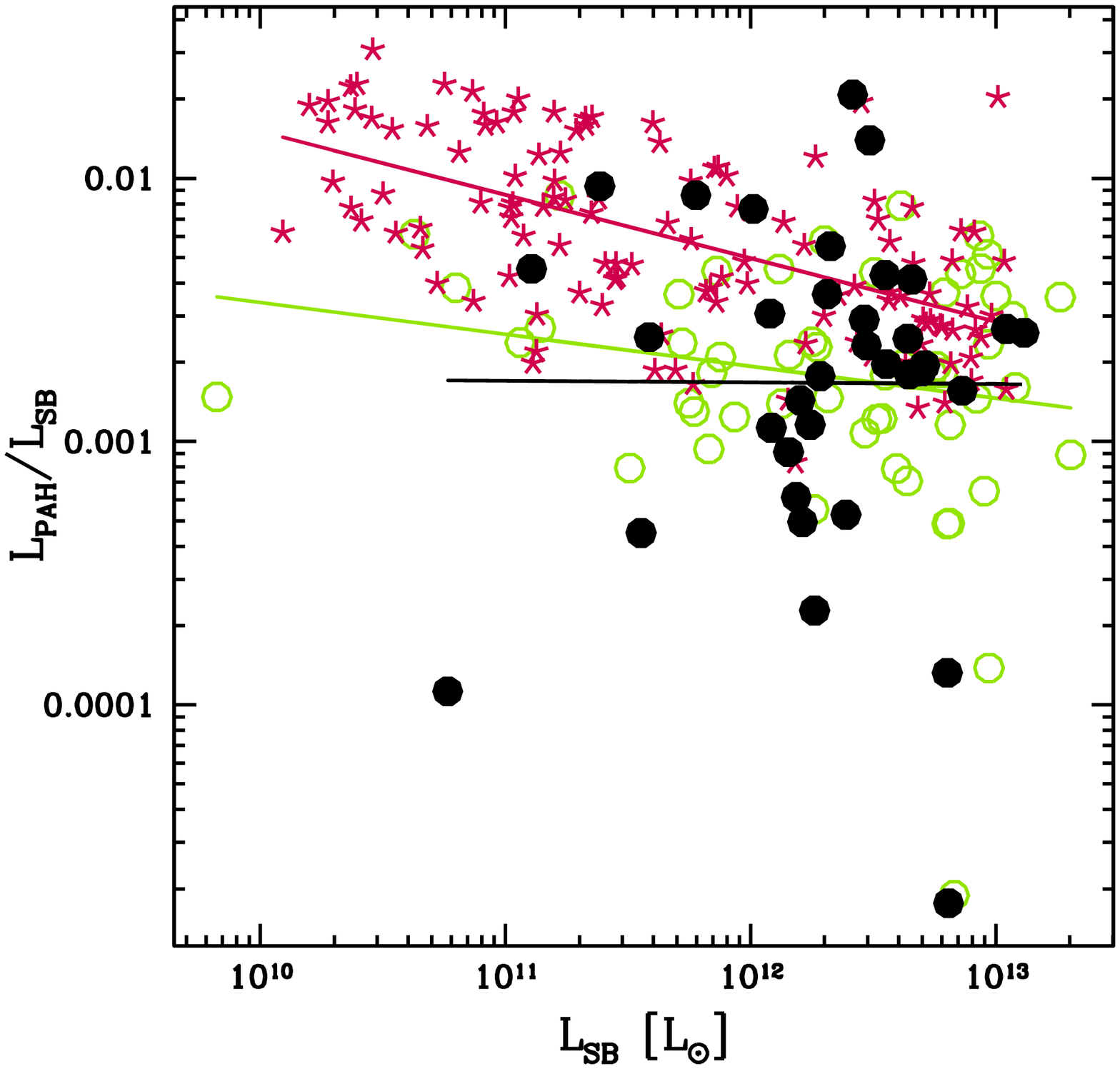}}
\caption{SFR$_{\rm FIR}$ versus SFR$_{\rm PAH}$ (top) and L$_{\rm PAH}$/L$_{\rm SB}$ as a function of L$_{\rm SB}$ (bottom) for unobscured AGN-dominated objects (filled circles), AGN-dominated objects with an optical stellar component (open circles) and starburst-dominated objects (stars). The continuous lines mark the respective linear correlations, the dashed line (top panel) the 1:1 relation.}
\label{fig:sfr}
\end{figure}

As other authors have previously pointed out \citep{schweitzer06,netzer07,lutz08}, SFR$_{\rm FIR}$ and SFR$_{\rm PAH}$ correlate with each other (Fig. \ref{fig:sfr}, top panel). Unobscured AGN-dominated objects (black circles) show a shallower slope and a loose correlation ($r=0.64$). AGN-dominated objects with an optical stellar component (open circles) show a tighter correlation ($r=0.82$) and a steeper slope that is almost parallel to the 1:1 line, but with systematically larger SFR$_{\rm FIR}$  than  SFR$_{\rm PAH}$ by about an order of magnitude. Finally, starburst-dominated objects (stars), have a very tight ($r=0.95$) correlation lying close to the 1:1 line at low SFR, with the SFR$_{\rm FIR}$ deviating from their SFR$_{\rm PAH}$ counterparts with increasing SFR. To exclude the possibility of these trends being a statistical effect, we ran the following test: we modelled a population lying on the 1:1 line with the scatter of the starburst-dominated objects and applied a random offset for the PAH luminosity of each object, drawn from a Gaussian distribution with $\sigma$ equal to the error on the PAH luminosity. We then recalculated the SFR$_{\rm PAH}$ and related EWs. The behaviour shown in the top panel of Fig. \ref{fig:sfr} was not reproduced and the AGN-dominated objects were distributed randomly, implying a physical origin of the observed SFR trend. The reason of this can be sought in the fact that the ratio of L$_{\rm PAH}$ to L$_{\rm SB}$ depends on L$_{\rm SB}$ \citep{smith07}. For example the ratio L$_{\rm PAH}$/L$_{\rm IR}$ has been found to be significantly smaller for local ULIRGs \citep{armus07} compared to that of regular local and low-luminosity starburst galaxies \citep{brandl06,smith07}. Moreover, high redshift ULIRGs, being on the whole less obscured than the local ones, present stronger PAH feature \citep{pope08,fadda10}.

\begin{figure}
\centerline{
\includegraphics[width=7cm]{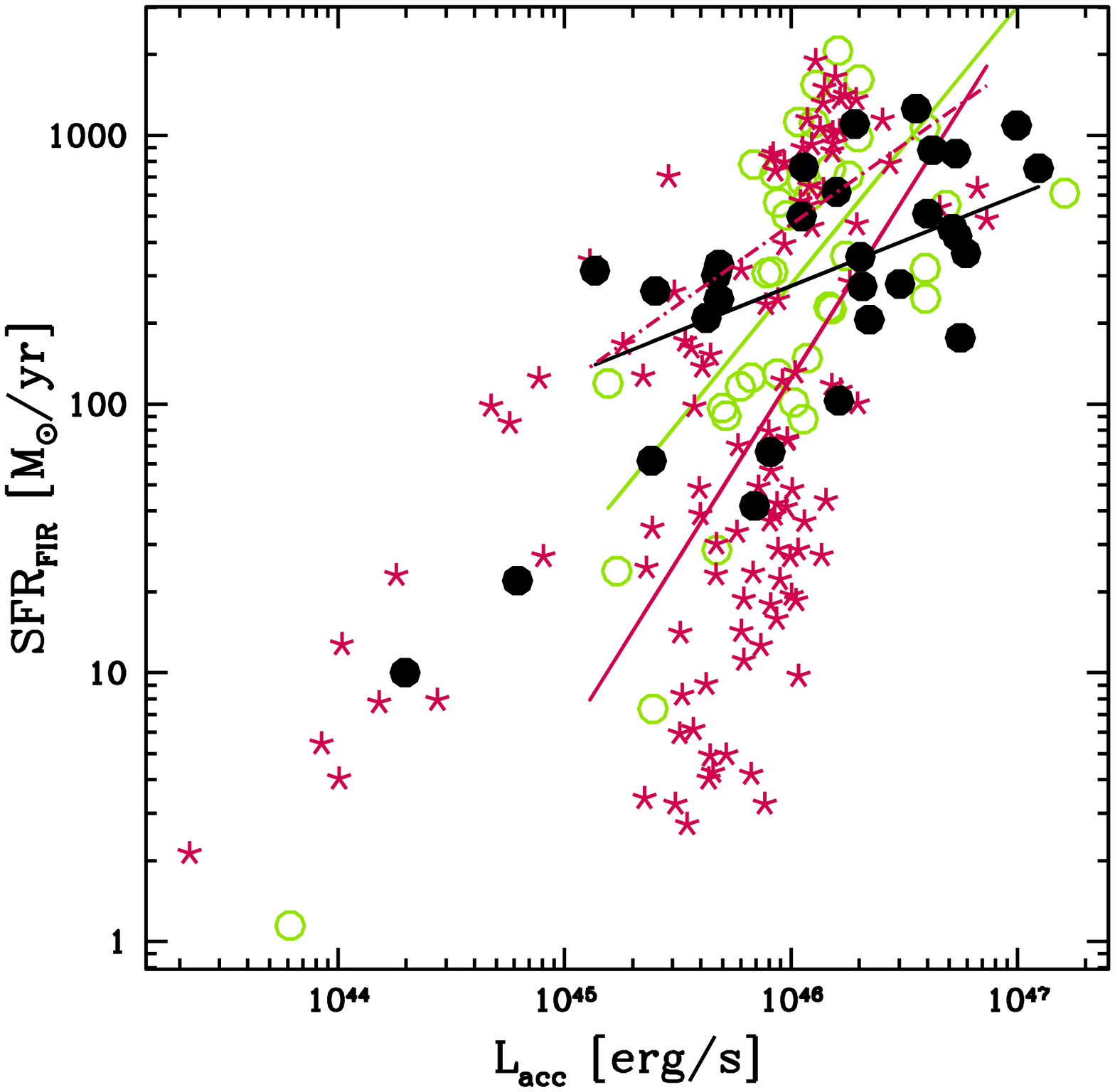}}
\centerline{
\includegraphics[width=7cm]{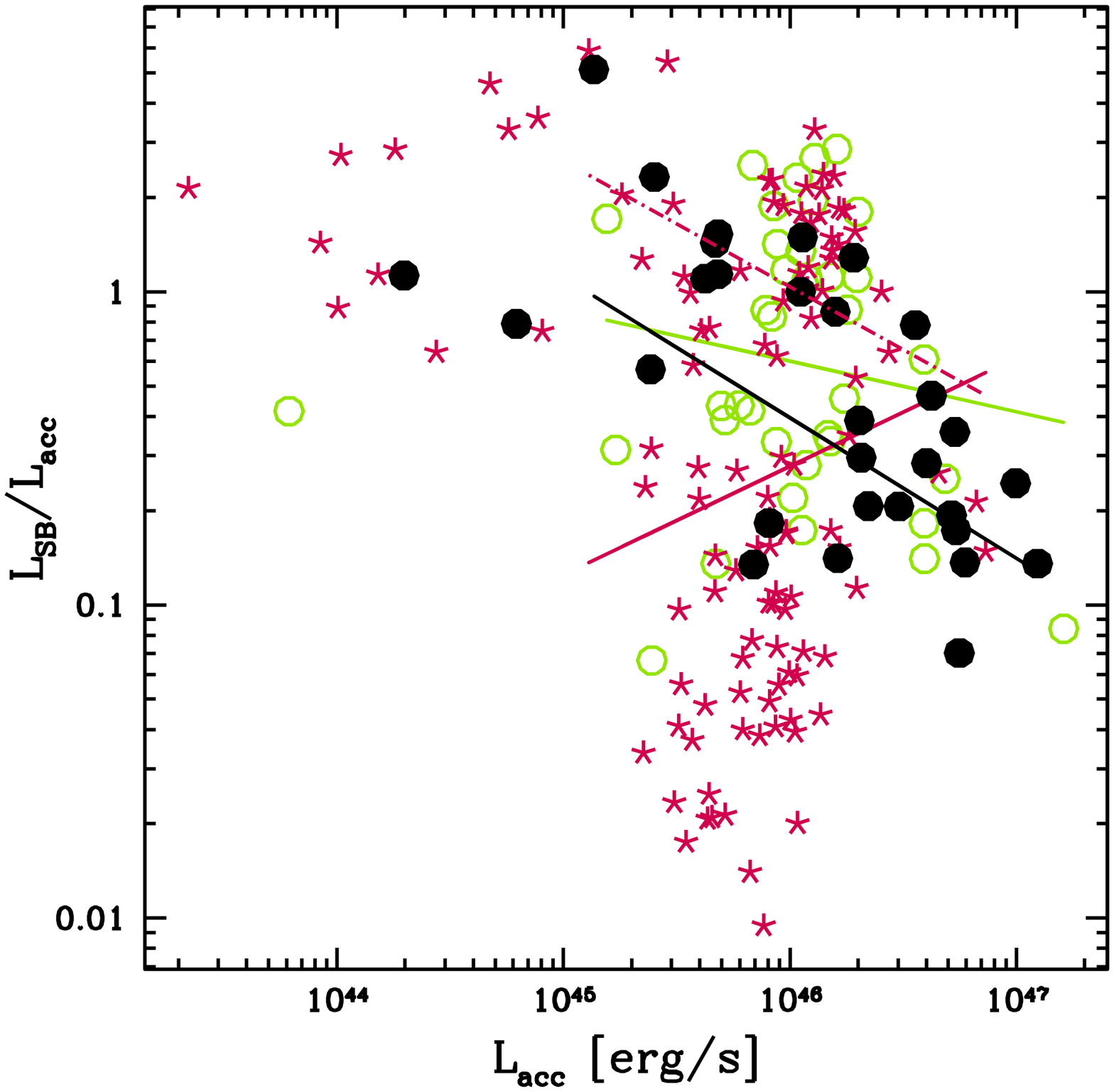}}
\caption{SFR$_{\rm FIR}$ versus L$_{\rm acc}$ and L$_{\rm SB}$/L$_{\rm acc}$ as a function of L$_{\rm acc}$ for unobscured AGN-dominated objects, AGN-dominated objects with an optical stellar component and starburst-dominated objects. The symbols-code is the same of Fig. \ref{fig:sfr}. The linear correlations shown here are computed for objects with L$_{\rm acc}>10^{45}$ erg/sec alone. The dotted-lines represent the correlation for starburst-dominated objects when discarding sources with N$_{\rm H}$ of the best torus model $< 3 \cdot 10^{23}$ cm$^{-2}$.}
\label{fig:sflacc}
\end{figure}

While comparing distant quasars with samples of local quasars and ULIRGs of \cite{schweitzer06} and \cite{netzer07}, \cite{lutz08} reported a constant L$_{\rm PAH}$/L$_{\rm SB}$ (and hence a constant  SFR$_{\rm FIR}$/SFR$_{\rm PAH}$, if we assume a constant SFR$_{\rm PAH}$/L$_{\rm PAH}$ ratio), with luminosities spanning almost four orders of magnitude. Analogously, \cite{pope08} found that sub-millimetre galaxies (SMGs) at z$\sim$ 2 allow to extend the relation between L$_{\rm IR}$ and L$_{\rm PAH}$ of the local starburst galaxies. In contrast, we find L$_{\rm PAH}$/L$_{\rm SB}$ to clearly decrease with L$_{\rm SB}$ for starburst-dominated objects (Fig. \ref{fig:sfr}, middle panel) while AGN-dominated objects have a close-to-constant L$_{\rm PAH}$/L$_{\rm SB}$, but with a large scatter. Furthermore, we find a decreasing L$_{\rm PAH}$/L$_{\rm SB}$ with increasing of redshift for starburst-dominated objects.

To see whether the presence of an AGN has an effect on the obscured SFR, we also looked at the behaviour of SFR$_{\rm FIR}$ as a function of the accretion luminosity, L$_{\rm acc}$, that is the normalisation of the AGN component to the observed data points (Fig. \ref{fig:sflacc}, top panel). As already seen in the recent literature, brighter AGN also exhibit higher SFRs \citep[e.g.][]{serjeant09,hatzimi10,serjeant10,bonfield11}. Focusing on the brightest objects of the sample, i.e. objects with L$_{\rm acc} > 10^{45}$ erg/sec, we see that unobscured AGN-dominated objects have a somewhat flatter distribution although the large scatter renders the correlation very weak ($r$=0.44). The increase of SFR seems to be less prominent for the very bright (L$_{\rm acc} > 10^{46}$ erg/sec) unobscured AGN but the low number of objects makes the derivation of any firm conclusion impossible. There is a tail of starburst-dominate objects with SFR$_{\rm FIR} <$ 100 and L$_{\rm acc} \sim 10^{46}$ erg/s for which the best fit torus models provide a quite high hydrogen column density, N$_{\rm H} > 3 \times 10^{23}$ cm$^{-2}$. This means that such objects are potentially heavily obscured in the optical band bringing more uncertainties in the estimation of the accretion luminosity. The dotted-line in both panels of Fig. \ref{fig:sflacc} represents the correlation of the starburst-dominated objects while excluding starburst-dominated sources with N$_{\rm H} > 3 \times 10^{23}$ cm$^{-2}$. The behavior of the starburst-dominated objects now resembles more that of AGN-dominated ones, albeit presenting a higher SFR. When comparing L$_{\rm SB}$ with L$_{\rm acc}$ we find their ratio decreasing with increasing L$_{\rm acc}$ for AGN-dominated objects but not for starburst-dominated ones. Analogously, when discarding starburst-dominated sources with high N$_{\rm H}$, the dotted-line in the bottom panel of Fig. \ref{fig:sflacc} is parallel to that of unobscured AGN, with the starburst-dominated objects having higher L$_{\rm SB}$ at same values of L$_{\rm acc}$.



\subsection{The cold and hot dust components}

The fit of the MIPS and SPIRE data with a grid of modified black bodies (as described in \S \ref{sec:bb}) returns the temperature of the cold dust component, heated by the starburst, as well as its mass. As already mentioned, we only fit a modified black body to objects with at least three data point at $\lambda > 100$ \mums, one of which has to be MIPS 160 \mums. The reason behind this choice is to sample both sides of the peak of the cold dust emission 
in order to avoid the introduction of possible biases \citep{shetty09a,shetty09b} and the under/over estimate of the cold dust temperature/mass \citep{smith12}. The mass of the hot dust is that of the torus models, defined as the integral of all dust grains over all volume elements.

\begin{figure}
\centerline{
\includegraphics[width=8.0cm]{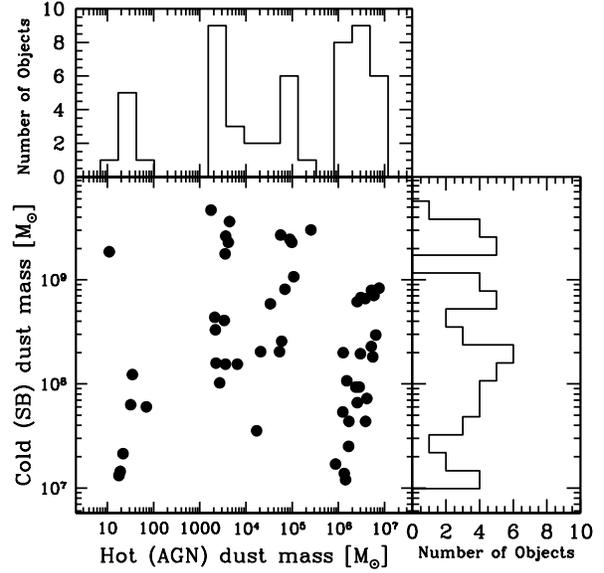}}
\caption{Cold (starburst-heated) dust mass versus hot (AGN-heated) dust mass.}
\label{fig:mass}
\end{figure}

The masses of the cold (starburst-heated) and hot (AGN-heated, that is the sum of all dust grains integrated over the torus volume elements) dust components do not correlate with each other, as seen in Fig. \ref{fig:mass}. These two components occupy very different physical scales: the hot dust, heated by the accretion disk surrounding the central engine, extends out to a few tens of pc, while the cold dust, heated mostly by young stars in star forming regions, extends to much larger distances from the central source, reaching kpc scales. Star formation is known to occur in individual or combinations of morphological features (such as spiral arms, central or extended starbursts, rings, etc) that are driven by gravitational instabilities, interactions and/or mergers. The lack of correlation between the masses of the hot and cold dust components therefore suggests that the gravitational effects that drive star formation do not divert a fixed fraction of the gas to the AGN center while the starburst is ongoing.

\begin{figure}
\centerline{
\includegraphics[width=7cm,angle=270]{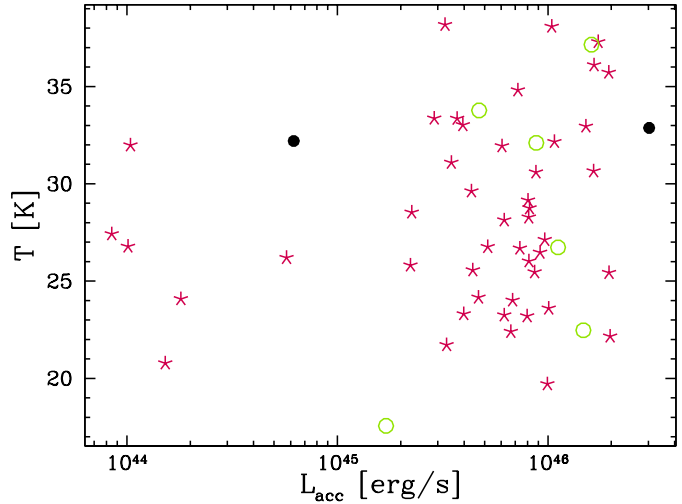}}
\caption{Cold dust temperature as a function of L$_{\rm acc}$ (circles and stars mark AGN- and starburst-dominated objects, respectively).}
\label{fig:tagnfrac}
\end{figure}

To check whether the presence of an AGN has an impact on the heating of the dust at large scales, we check the correlation between the derived black body temperatures and the AGN accretion luminosity, L$_{\rm acc}$, as shown in Fig. \ref{fig:tagnfrac}. Unfortunately, the requirement for a 160 \mum detection to constrain the shape of the SED at $\lambda < 200$ \mum limits the number of objects for which the temperature of the cold dust can be determined as the 160 \mum data are very shallow: this affects many AGN-dominated objects (open and filled circles). Due to the small number of available data points we used a single-temperature modified black body component to account for the emission at FIR wavelengths where both warm and cold dust can contribute. The range of temperatures, reported in Table \ref{tab:laccT}, that we fit is consistent with temperatures found by \cite[e.g.][]{kirkpatrick12} using a multi-temperature modified black body approach that considers two modified black bodies to account for the warm and cold dust components: the majority of the temperatures derived for the cold dust component span the range between 20 and 40 K, with a mean temperature of 28.5 K. Even though the majority of objects shown in this figure are starburst-dominated in the MIR, many harbour an AGN with high L$_{\rm acc}$ values. With the above caveats in mind, we find no evidence that the cold dust temperature is affected by the presence of an underlying AGN. This can also be seen from the average temperature of the different L$_{\rm acc}$ bins (reported in in Table \ref{tab:laccT}).

\begin{table}
\small
\begin{center}
\begin{tabular}{cccccc}
\hline
L$_{\rm acc} [erg/s]$ & \# & $\langle$T$\rangle [K]$  & $\sigma_{\rm T}$ [K] & T$_{\rm min}$ [K] & T$_{\rm max}$ [K]\\
\hline
total & 53 & 28.5 & 5 & 17.6 & 38.2 \\
$\leq 10^{45} $ & 7 & 27.1 & 3.8 & 20.8 & 32.2 \\
$10^{45} - 10^{46}$ & 33 &  27.7 & 4.6 & 17.6 & 38.2 \\
$> 10^{46}$ & 14 & 30.9 & 5.6 & 22.2 & 38.1 \\
\hline
\end{tabular}
\end{center}
\caption{Results of the temperature of the cold dust both for all the objects with MIPS 160 and for different bins of L$_{\rm acc}$. From the left-most to the right-most column: range of L$\rm {acc}$, number of sources, mean temperature, $\langle$T$\rangle$, standard deviation, $\sigma_{\rm T}$, and minimum and maximum temperature, T$_{\rm min}$ and T$_{\rm max}$, respectively.}
\label{tab:laccT}
\end{table}


\section{Conclusions}\label{sec:conclusions}

Assessing the effects of the presence of an active nucleus in the centre of a galaxy is of paramount importance to the understanding of the evolution of the galaxy and the coevolution of the activity processes occurring during the galaxy's lifetime. 
In this paper we present the analysis of a sample of 375 extragalactic sources in the northern HerMES fields of Bootes, FLS, Lockman and ELAIS N1 with available broad-band photometry spanning the optical (SDSS) to the FIR ({\it Herschel}), and {\it Spitzer}/IRS spectra, with the aim to investigate the observational signatures of AGN in the MIR and FIR wavelengths and their impact on the properties of their hosts. The IRS spectra impose constraints on the AGN torus models while the SPIRE photometry is essential for the measurement of the cold dust properties of the host galaxies. Spectrophotometric multi-band and multi-component SED fitting, in combination with EWs and luminosities of the PAH features measured from the IRS spectra, allows us to investigate the source properties as a function of AGN content.  


We find SFR$_{\rm FIR}$, the obscured star formation rate derived from the IR luminosity of the starburst component, and SFR$_{\rm PAH}$, derived from the luminosity of the PAH features, to correlate with MIR AGN- and starburst-dominated populations presenting different correlations. Moreover, we note that, as a general trend, the SFR$_{\rm FIR}$ takes systematically higher values than SFR$_{\rm PAH}$, with the possibility of this being due to statistical errors excluded. We find L$_{\rm PAH}$/L$_{\rm SB}$ to be almost constant for AGN-dominated objects but to decrease with increasing L$_{\rm SB}$ for starburst-dominated objects, contrary to what has been reported by other authors \citep[][and references therein]{lutz08}. Furthermore, we observe an increase in SFR with increasing L$_{\rm acc}$, with the increase less prominent for the very bright, unobscured AGN-dominated sources. 

We find no noticeable effect of the presence of an AGN on the FIR properties of the host galaxy: SFR$_{\rm FIR}$ increases with increasing L$_{\rm acc}$, as already reported in the recent literature \citep[e.g.][]{serjeant09,hatzimi10,serjeant10,bonfield11}. We find the ratio L$_{\rm SB}$/L$_{\rm acc}$ to decrease with increasing of L$_{\rm acc}$. No significant dependence of the temperature of the cold dust on L$_{\rm acc}$ is observed, even for sources harbouring bright AGN components with L$_{\rm acc}>10^{46}$ erg/sec. 

The cold dust, heated by young stars in star forming regions, extends to several kpc from the central source. The hot dust emission, on the other hand, arises from the pc-size region surrounding the central black hole and could, in principle, serve as the reservoir that feeds the black hole during the high energy accretion phase of a galaxy. We find the masses of the two dust components to be completely uncorrelated and interpret this lack of correlation as an indication that the fraction of gas funnelled to the AGN as a result of gravitational effects that also drive the starburst activity is not constant.

To summarize, our findings are in agreement with there being no evidence for the AGN significantly influencing star formation processes of the host galaxy. This is consistent with the fact that most models predict an extremely brief feedback phase: when considering large IR samples, an average effect is expected to be observed. This implies that a correlation between the hot and cold dust properties is not expected to be seen, even with a feedback itself being very strong. 

Our findings are based on spectral synthesis techniques that are methodology-dependent, and the models considered not completely free from degeneracy. Nonetheless, they do not support the scenario in which the AGN in a galaxy's center has an impact on the star formation of the host \citep[as also found by e.g.][]{hatzimi10,harrison12,santini12,rosario12} but show, instead, that the two phenomena coexist in a variety of both AGN- and starburst-dominated sources spanning more than four orders of magnitude in both L$_{\rm acc}$ and L$_{\rm IR}$. 


\section*{ACKNOWLEDGMENTS}
 
SPIRE has been developed by a consortium of institutes led by Cardiff Univ. (UK) and including: Univ. Lethbridge (Canada); NAOC (China); CEA, LAM (France); IFSI, Univ. Padua (Italy); IAC (Spain); Stockholm Observatory (Sweden); Imperial College London, RAL, UCL-MSSL, UKATC, Univ. Sussex (UK); and Caltech, JPL, NHSC, Univ. Colorado (USA). This development has been supported by national funding agencies: CSA (Canada); NAOC (China); CEA, CNES, CNRS (France); ASI (Italy); MCINN (Spain); SNSB (Sweden); STFC, UKSA (UK); and NASA (USA). 

The SPIRE data presented in this paper are available through the HerMES Database in Marseille (HeDaM; http://hedam.oamp.fr/HerMES).

The Cornell Atlas of Spitzer/IRS Sources (CASSIS) is a product of the Infrared Science Center at Cornell University, supported by NASA and JPL.

Funding for the creation and distribution of the SDSS Archive has been provided
by the Alfred P. Sloan Foundation, the Participating Institutions, the National
Aeronautics and Space Administration, the National Science Foundation, the U.S.
Department of Energy, the Japanese Monbukagakusho, and the Max Planck Society.
The SDSS Web site is http://www.sdss.org/.

Much of the analysis presented in this work was done with TOPCAT (http://www.star.bris.ac.uk/$\sim$mbt/topcat/), developed by M. Taylor.

\label{lastpage}

\end{document}